Fusion breeding and pure fusion development – perceptions and misperceptions

Wallace Manheimer
Retired from the US Naval Research Laboratory,
wallymanheimer@yahoo.com
Orcid number:  0000-0001-6334-2591

Abstract

This paper examines fusion breeding, namely the use of 14 MeV fusion neutrons to breed $^{233}$U fuel for thermal nuclear reactors.  This can be accomplished much more quickly than pure fusion.   It can become main component of a power architecture that is economical, enviromentally sound  and has little if any proliferation risk.

I.      Introduction

This author believes that fusion is one of the few possible methods of supplying sustainable energy for future civilization in an economically and environmentally acceptable way.  Traditionally, the development of pure fusion, as opposed to hybrid fusion or fusion breeding, has path has been government support leading to implementation in the economy in 35 years, or as skeptics say, 35 years away and always will be.

More recently there has been a perception of an imminent climate disaster, motivating a rush on the part of many in the in the fusion community, publicly and privately financed,  to attempt to implement fusion on a much faster time, basically within a decade.  There have been many new companies,  'fusion start ups', now promising fusion in a decade, motivated in part as a response to the supposed imminent 'climate disaster'.

Which of these paths is the right one?  This author's answer is neither one.  He makes the case that the the development time for pure fusion will not only be much more than a decade, not only much more than 35 years, but is extremely unlikely to be in this century.   Furthermore, this paper argues that there is no oncoming climate disaster. We should not exploit the fear of a false disaster to temporarily advance the funding for an impossibly rapid development of fusion. When these rapid efforts fail, the harm done to the fusion project could be immeasurable.

All these privately funded ' fusion start ups' will fail and the investors will most likely lose their investment. However there is an alternative which could be ready in the often named 35years. This is using a fusion reactor to breed conventional nuclear fuel for thermal nuclear reactors. As a breeder, the requirements on the fusion reactor are relaxed by at least an order of magnitude.

The options for sustainable power are few. Regarding the role of windmills and solar panels, there is a large literature, provided by undisputed experts, and backed up by bitter experience, showing that are not only are they extraordinarily expensive, unreliable, and wasteful, but far from being the cleanest power source, they are the dirtiest [1-12]. The amount of material needed, land needed, and difficulty of disposal at the end of useful life makes them a much, much greater source of pollution and destruction of land than does any other power source. Furthermore, this attempted replacement would be especially tragic if the effort were futile; the cost, astronomical; and the necessity, nonexistent.

If windmills and solar cannot provide for economically and environmentally acceptable, reliable sustainable power, and fossil fuel and mined $^{235}$U are limited, non sustainable resources, and pure fusion a very, very distance possibility, what else is there besides breeding fissile material from fertile material? There are breeding possibilities from fission of course. India and Russia are actively looking into these, and Russia has two fast neutron breeders, their BN 600 and BN 800 currently hooked up to their grid (the Russian word for fast is bistro). However it is also possible to breed via fusion, and although the fusion breeding development path is obviously longer, fusion breeding has major undeniable advantages over fission. Breeding nuclear fuel, by either fission or fusion opens up a resource that can supply civilization at 40 terawatts at least as far into the future, as the dawn of civilization was in the past.

Accordingly there are misperceptions, and perceptions on the best path for fusion. This author's responses to the misperceptions: The climate has been and always will be changing, but there is no climate crisis. Solar and wind are totally inadequate solutions to this non crisis. The fusion start ups will all fail, and the will be no commercial fusion in 10 years, in 35 years, or in the 21st century.

Regarding the proper perceptions: Fusion breeding along either the ITER (International Tokamak Experimental Reactor) or NIF (National Ignition Facility) pathways may well be able to provide nuclear fuel not too long after midcentury, but cannot play a role in the world energy budget as pure fusion devices at least in the $21^{st}$ century. Fusion breeding could be the key development to make 'energy parks' worldwide by combining the best aspects of fusion and fission energy.

We will deal with climate, only briefly here, but there is voluminous documentation of the claims made here that there is no climate crisis [3,5,13-20]. Perhaps the best statement summarizing the skeptics case was made by Ricahrd Lindzen, most likely the worlds foremost authority on geological fluid motion:

"What historians will definitely wonder about in future centuries is how deeply flawed logic, obscured by shrewd and unrelenting propaganda, actually enabled a coalition of powerful special interests to convince nearly everyone in the world that CO2 from human industry was a dangerous, planet-destroying toxin. It will be remembered as the greatest mass delusion in the history of the world- that CO2, the life of plants, was considered for a time to be a deadly poison."

Regarding special interests, the skeptics are often accused of being shills for oil and coal companies, but the reality is that the hundreds of billions, no trillions that has gone to the support of climate alarmism supports what Bjorn Lomborg has called 'The climate industrial complex' (13). It begins:

The tight relationship between the groups echoes the relationship among weapons makers, researchers and the U.S. military during the Cold War. President Dwight Eisenhower famously warned about the might of the "military-industrial complex," cautioning that "the potential for the disastrous rise of misplaced power exists and will persist." He worried that "there is a recurring temptation to feel that some spectacular and costly action could become the miraculous solution to all current difficulties."

This is certainly true of climate change. We are told that very expensive carbon regulations are the only way to respond to global warming, despite ample evidence that this approach does not pass a basic cost-benefit test. We must ask whether a "climate-industrial complex" is emerging, pressing taxpayers to fork over money to please those who stand to gain.

And concluding with:

The partnership among self-interested businesses, grandstanding politicians and alarmist campaigners truly is an unholy alliance. The climate-industrial complex does not promote discussion on how to overcome this challenge in a way that will be best for everybody. We should not be surprised or impressed that those who stand to make a profit are among the loudest calling for politicians to act.

Actually, one does not have to be a climate expert to contribute to the skepticism. This author, a scientist with more than 50 years of experience, but not as a climate scientist, has played minor role. One can simply check out the predictions of the prophets of doom against a Google search, and see that for the most part these are grossly exaggerated [21]. Anyone can do this anywhere, any time, there is no need for a 'climate scientist' to guide us. Furthermore, one can do the same with media reports and conclude that most of the legacy media plays a very one-sided role in its reporting on the climate dilemma [22]. Full disclosure: other scientists have criticized this work [23]. The author's response to their criticisms is recorded in Ref. (24). Actually, Reference (24) lists a large number of climate experts who dispute the standard dogma of a climate emergency. While it is a subjective matter, to this author, his list is one of scientists who have far more knowledge, experience, and gravitas regarding the climate dilemma than do most of the scientists who have made a name for themselves by predicting climate gloom and doom. Finally, most of the projections of calamity are made by running numerical simulations of the climate. This is an area in which the author does have considerable experience in his 50+ year career and has pointed out some of the pitfalls of this approach [25] as have many others. [3,26,27]

As a final indication of the lack of confidence that the threat of a climate crisis is real, there was a large international meeting to discuss the climate dilemma in Scotland in November 2021. World leaders, including President Biden and many European leaders attended. However, the leaders of Brazil, Russia, China and Turkey voted with their feet, and did not attend. The leader of India attended but announced that India would not be reducing its CO2 emission until 2070, an absolutely meaningless commitment. These are large, important, technically advanced countries, containing ~ 40% of the world's population.

Actually, the western democracies are not all that different. Typically, some bureaucrat orders that we have to stop or reduce the use of fossil fuel in this way and that. Occasionally the new rule is put to a vote, and the new rule is almost always rejected by the voters. As Yogi Berra put it "If people don't want to come to the ballpark, you can't stop 'em"

Modern civilization does need energy. Without energy other than human and animal muscle, civilization would be a thin veneer, atop a vast mountain of human squalor and misery, as it has been for almost all human history. This added energy over the last ~ 200 years has allowed the benefits of civilization spread to billions of additional people. When people or institutions insist that we drastically change our energy system, by disassembling what we have now and installing something new; one must realize that if this new system fails, as solar and wind almost certainly will; it would mean the end of modern civilization. The stakes in the battle are immense.

Hence one misperception is that fusion is a cure for the supposed climate crisis and must be implemented quickly to stave off disaster. This author not only disbelieves this, but believes that such a path could well be ultimately be very harmful for the development of fusion, and perhaps for future civilization itself. He believes that there is plenty of fossil fuel for 35-50 years and using it will not produce any climate crisis. However it is not a sustainable path for world development, and the author believes that if prosperous civilization is realized world wide, fossil fuel is a much less sustainable resource than current estimates. This then is the necessary time scale for the development of fusion breeding, and the author believes that indeed this is possible with a properly focused fusion program.

Fusion breeding might well be achievable not too long after midcentury, and could likely be needed then. This author has written several articles, advocating fusion breeding [11,28-38], including several open access articles [11,34-37] in prestigious journals, and has written one simplified version in the APS Forum on Physics and Society. [38] However, fusion breeding has been the ugly duckling of the fusion project, rejected by most, but certainly not all, of both the fission and fusion community. But it should no longer be ignored, nor ignorantly condemned with such erroneous remarks as: "it combines the worst aspects of fusion with the worst aspects of fission", or "fusion breeding can only address fuel, the one problem that fission does NOT have". This paper, and others, hope turn fusion breeding into the beautiful swan. In fact, it has much, much greater potential than simply not being written off. This article makes the case that not only *can* it breed, it is the *best* breeder.

This paper is an attempt to determine the best way to develop and implement fusion breeding into the economy. It proposes a very different optimum strategy than what is usually proposed. Section II uses a 'top down', rather than 'bottom up 'analysis to determine the power necessary to support modern civilization world wide. It makes the case that fusion breeding will be necessary by about midcentury and very likely can be developed by then or not too long thereafter. Section III claims that the fusion 'startup' companies promising fusion hooked up to the grid in 10 years or so will all fail. Section IV argues that fusion on the normal government sponsored path will not give rise to fusion hooked up to the grid in this century. Section V makes the case that fusion breeding, that is the use of 14 MeV fusion neutrons to breed $^{233}$U as fuel for thermal nuclear reactors, could be a sustainable approach for powering civilization not too long after midcentury. It will not be needed before that, but could be desperately needed afterwards. Section VI suggests an optimum strategy to achieve fusion breeding, by magnetic or inertial fusion, well before century's end. Section VII shows that even if miraculously, a pure fusion machine becomes available by midcentury, breeding may still be the best option. Section VIII describes the 'energy park', a collection of one fusion breeder, one fast neutron reactor and 5-10 thermal nuclear reactors that could be the basis of an infrastructure that sustainably powers civilization.

To summarize, the proper role for fusion (i.e. fusion breeding) is to prepare a sustainable energy source, ready to preserve the benefits of civilization, worldwide, by midcentury or not too long thereafter. It is *not* to be developed in the next decade to solve the nonexistent climate crisis. This latter attempt would be an enormous waste of precious fusion resources. Even with its large private dollar support, the false hope it engenders would be much more likely to harm the fusion enterprise than to benefit it.

II. Sustainable Energy from the top down

The goal is to increase the world's energy so everybody has a life style like the more economically advanced nations, and to do it by mid century or not too long after. This means examining the energy development requirements from the top down, not from the bottom up, as we explain shortly.

One excellent source of these statistics is the yearly publications by BP. [39] Taken from their 2019 issue are their graphs of the sources of energy, the energy use in various parts of the world, and by end use.

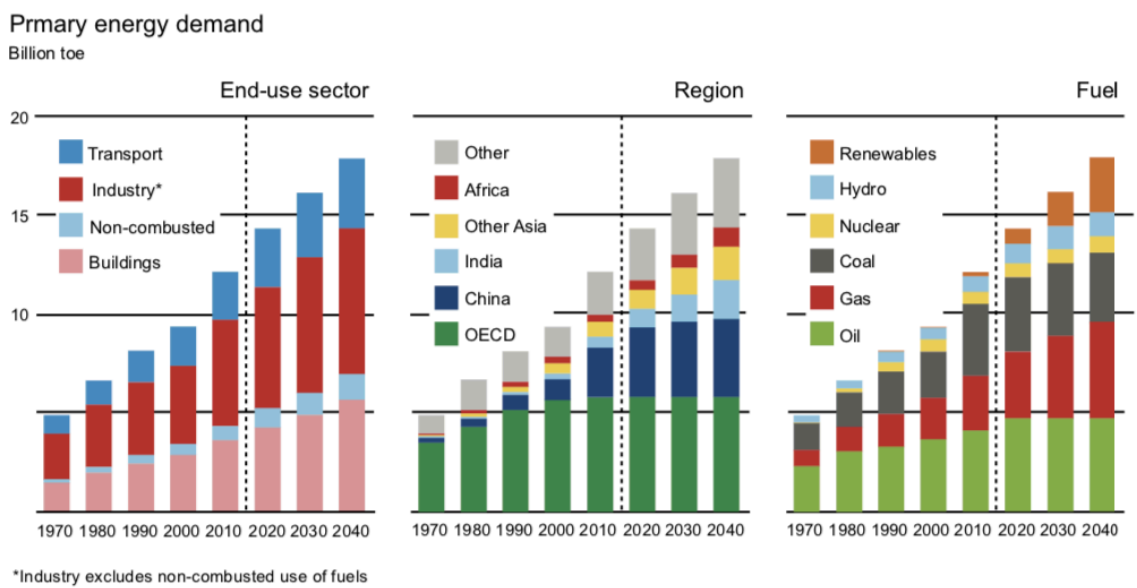

Figure (1):  BP's three graphs of energy demand by end use sector, region and fuel.  The units are billions of tons of oil per year.  Since this is an unusual unit for people not in the oil industry, we use terawatts (TW), where one terawatt is approximately one billion tons of oil per year.

To the left of their vertical dashed line in Fig. (1) is the historical record. To the right are BP's extrapolations for the future.  While up to 2040 BP sees fossil fuel as playing a very important role, it is not sustainable.  That is it will run out at some point.  As is apparent from the graph, the world today uses about 14 terawatts (TW).  However the energy use is very unequal.  The 1.2 billion people in the economically more advanced OECD countries use ~ 6 TW, or ~ 5 kilowatt (kW) per capita.  The other 7 billion or so people living on the planet share 8 TW, or use ~ 1 kW per capita.  The world's goal certainly must be to bring the rest of the world up to OECD standards of living as quickly as possible. By mid century the world population is expected to level off at about 10B, meaning that at current OECD power use, the world would need 50TW!

However energy efficiency is also expected to increase, and typically increases by ~ ½ - 1% per year. [40] Hence let us think in terms of 35-40 TW by mid century. Two ways the energy efficiency could increase are to switch from conventional coal powered generators at ~33% efficiency, to more efficient ultra super critical coal powered generators at ~ 40%. These generators also have advanced pollution control and emit only water and carbon dioxide. [41] Furthermore, a portion of electricity generation could shift to gas powered generators, which typically run at ~ 50% efficiency. Undoubtedly there are many other approaches to increasing efficiency, including having more business meetings on line instead of in person.

The BP extrapolation up to 2040 is a perfect example of extrapolation 'from the bottom up'. They take what is being done now, and and see how much they can push it to advance. They see a power increase of up to ~ 2 TW per decade by 2040, or an increase of ~ 200 GW per year. Extrapolating their graph to 2050, they would probably predict a world wide power of ~20TW. However this is not nearly sufficient to bring the world up to OEDC standards by then.

This author prefers interpolation 'from the top down'. That is, let's see how much power we need to get the world up to OECD standard at some particular time, which we take as mid century, 2050. This means increasing the power by ~ 21 - 25 TW by then, or ~ 7-8 TW per decade; more than triple the advance that BP foresees! This should be the goal the world strives toward.

Whether the concern is exhausting fossil fuel (we can use it for quite a while, but will exhaust it in 1/3 the time at 35TW as at 12), or is knowing that solar and wind cannot do the job; [1-12] or knowing that pure fusion cannot do the job, at least in this century if ever (see the next two sections) these lead to one and only one conclusion. Nuclear power must play an important role, both in any final sustainable role, and on the way there. Let us think of a sustainable future for all mankind as one that increases nuclear power by about a factor of 20 to ~ 20TW (i.e. ~7TWe) worldwide by midcentury, reducing fossil fuel somewhat to ~10TW, so it will last at least as long as current estimates, and increasing hydro and renewables to 1-3 TW each. In other words, it still recognizes that fossil fuel will play an essential role, but less crucial than today. It does not regard the use of fossil fuel, at 10 TW well into the future, as causing an extreme planetary calamity.  Ultimately, as fossil fuel runs out, nuclear power would take over completely, but would be quite far in the future.

This then would obviously require something of a crash program in expanding nuclear power over the next few decades.  There is every reason to think this possible technically, although perhaps not politically.  At least in the United States, regulations, lawsuits, protest marches, bureaucratic delays, environmental impact statements done and redone numerous times, NIMBY, BANANA,… have all thrown sand in the gears of nuclear power. These could be the biggest problem it faces.  Even if the nuclear company is successful, typically 20 years are wasted as it strangles in bureaucratic red tape and court cases, enormously increasing the price of nuclear power.  Regulation reform is the American, and perhaps the worldwide nuclear industry's biggest battle right now.

It may be that nuclear power is making a comeback.  John Kerry, the man most responsible for killing the American breeder program in 1994 is now saying "Go for it" regarding a renewed American nuclear program.   Prime minister Boris Johnson, before his resignation (i.e. in May 2022) said the UK will build one new nuclear plant a year.  Also, it is planning an advanced breeder (an upgrade to the American Integral Fast Reactor, IFR) called the PRISM to treat its plutonium wastes.  In February 2022 France announce plans to build at least 6 new reactors, and perhaps even an additional 8.

Yet even if the nuclear industry solves its image problem, it faces a much bigger problem on the physics and technical side.  Fissile $^{235}$U comprises only 0.7% of the uranium resource.  Supplies of mined 235U are limited, almost certainly much less than the reserve of fossil fuel, and uranium dissolved in the sea are almost certainly at much too low a concentration to have any impact. [42]  One rather pessimistic estimate is that the energy resource of mined uranium is about 60-300 Terawatt years. [30]  Other estimates are higher, but no estimate is high enough, that if it were correct, there would be enough uranium to sustainably supply the world's thermal nuclear reactors with 20-30TW (i.e.  ~6-10TWe).

Ultimately breeding fuel must play an important role.  Breeding means taking a material which exists in nature like, $^{238}$U or $^{232}$Th, called fertile materials, and bombarding them with neutrons to make fissile materials, like $^{239}$Pu or $^{233}$U, which do not exist in nature. However, because these have an odd atomic weight, they are fine as fuel for thermal fission reactors such a light water reactor (LWR). There are certainly conventional approaches to breeding, including fast neutron reactors, [43-46] and thermal thorium reactors, [47] and these certainly have a shorter development path than fusion breeding. Other countries, especially Russia and India are taking these reactors very seriously. Russia already has two fast neutron reactors, their BN 600 and BN 800 attached to their grid. However, fission breeders are complicated and expensive.

Not only is the reaction cross section much greater for a thermal neutron reactor, but the thermal reactor designer has a wide choice of coolants (e.g. water or air), instead of only liquid sodium or lead, which must be used in a fast neutron reactor. Figure (2) is a plot of the fission and neutron absorption cross sections as a function of neutron energy for $^{235}$U and $^{238}$U. [48]

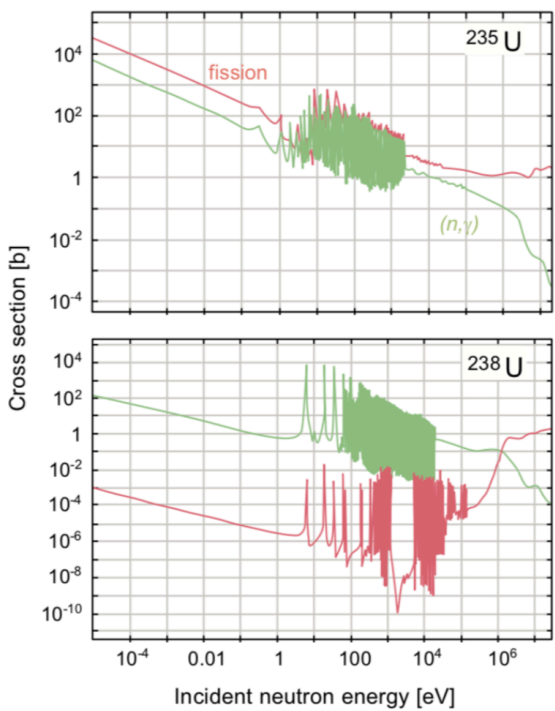

Figure 2 : The fission and neutron absorption cross section in barns (1 barn is $10^{-24}$ cm) for $^{235}$U and $^{238}$U as a function of the energy of the incident neutron. The cross sections look about the same for all fertile and fissile nuclei, depending whether their atomic number is odd or even.

> The red curves are the fission cross sections, and the green, are the neutron absorption cross sections.

While fusion breeding certainly has a longer development path than fission breeding, it does have important advantages which we discuss shortly.

III: Fusion start-ups' pilot plants all have serious problems and will most likely fail

In the last few years, many private companies, generally called 'fusion start ups', have sprung up. They claim, nearly universally, that they will have fusion on the grid much quicker than the national labs, generally within about a decade. They have achieved wide recognition, and in March 2022 there was even a White House conference on their approach to fusion. At the conference, the fact that rapid development of fusion would solve the climate crisis played center stage. Here is a portion of Gina McCarthy's talk while hosting the conference[49]:

Gina McCarthy, White House National Climate Advisor, spoke about the ways the Biden-Harris Administration is addressing climate change, such as a commitment to get to 100% carbon-free electricity by …..

Needless to say, the various 'fusion start ups' snapped at the bait. Here are two quotes typical of very many:

General Fusion web site: With the urgency of climate change in mind, we are on course to power homes, businesses, and industry with fusion energy by the 2030s.

TAE press release: "We're here today because the world is on fire. Because generations before us made bold investments to give us the first solutions to this crisis. Because we must act. And we believe that when it comes to fusion, the time is now."

This paper asserts that thinking of fusion hooked up to the grid in ~10 years is a pipe dream. The effort is simply not ready for prime time, there is too much unknown to get from where we are now, to electricity for the grid in the time period these companies promise. In fact, given the lack of a 'climate crisis', they are peddling a non-solution to a non-problem, this nearly immediate non-solution, which will fail, will cost billions.

This author is certainly not opposed to private companies ultimately getting involved in fusion once the time is right. In fact, ultimately, they will be essential, but the time is not now. Consider the situation in space capable rockets, where private companies in the United States are now eclipsing NASA rockets (in fact originally Army and Navy rockets). But NASA rockets were the only game in town for about 50 years, and NASA made enormous gains in developing these rockets with no input from the private sector for about half a century. Only then, when much of the technology was largely established by NASA, could the prospect of further development be more certain, and in a shorter time, so that investors could reap profits within a reasonable time horizon if the project were to be successful. Only then could private companies come in and begin to take over. Fusion is nowhere near where space rockets were even as early as in 1957 when satellites began to circle the earth. After all, rocket work began decades before 1957. We have not even launched our fusion analog of Sputnik yet. Private companies could not take over space rockets in 1956; private companies cannot take over fusion now.

These companies are supported by private dollars, and the dollars invested are large. Helion brags that it has attracted $2.2B; Commonwealth Fusion, $1.8B; TAE, ~ $1B. These companies, of course, can spend their private $$$ any way they wish, but if their investors expect a payoff in the next decade, they are in for a big disappointment.

While with their large dollar amount, they will undoubtedly make some contributions to fusion science and technology, this author thinks their overall impacts will be harmful. They are raising unrealistic expectations, and who knows what the effect will be when these efforts all collapse with a gigantic splat. Will they take down the normal, government sponsored fusion effort with them, thereby destroying a potentially very important project? Quite possibly.



Why will these efforts most likely fail?

Commonwealth Fusion: This company, a spin off from the MIT program, is probably the only one that could have succeeded in going the normal government sponsored route. They have published their designs in the open literature,[50,51] so it is entirely appropriate for readers to offer scientific analysis of their design, and if appropriate criticism, also in the open scientific literature. They have a new technology that could allow tokamaks to operate at larger magnetic fields, 9-10T instead of 3-5T. The Princeton Plasma Physics Lab (PPPL) at one point was interested in purchasing these field coils to do their own experiments with them. [52] Commonwealth has plans for a small tokamak called SPARC, which they hope demonstrates high gain, and a larger one called ARC which they will hope will deliver commercial power in a decade or so. The company has been rather open and has published a considerable amount of its work.

Regarding the plasma, the higher magnetic fields are an unquestioned benefit, but in reality, they are a double edge sword. An economical fusion source is more than just the plasma, and for these other components, the high field is a drawback. As the size shrinks and the power increases, the wall loading increases. SPARC and ARC would have higher wall loading than ITER, [53,54] and wall loading is a serious problem area for ITER. Not only that, unlike ARC, ITER is a pulsed machine, and does not have to operate steady state so as to supply power to the grid. Table I gives the radius in meters, field in T, the power and the neutron wall loading for ITER, SPARC, and ARC, taken from the ITER web site, as well as the Commonwealth papers just cited.

| Machine | R (M) | B(T) | P(MW) | loading (MW/m$^2$) |
|---|---|---|---|---|
| ITER | 6 | 5 | 500 | 0.7 (pulsed) |
| SPARC | 1.65 | >10 | 50 | 0.8 (steady state) |
| ARC | 3.3 | 9 | 500 | 2.3 (steady state) |



TABLE I, The major radius, magnetic field, fusion power, and neutron wall loading, as expected for several tokamaks.

Notice that SPARC will have more wall loading than ITER, and ARC, considerably more. So far, no tokamak, or any other fusion reactor, has operated with any neutron wall loading at all. To think that ARC and SPARC can operate steady state with higher wall loading than a pulsed ITER, is optimistic, to say the least.

Perhaps even more important, neither ITER, nor SPARC, nor ARC knows at this point how it will drive current steady state. ITER, and most tokamaks drive the current with a transformer, the plasma being the secondary. But the transformer can only drive so many Volt seconds and then the current stops. ITER's pulse time will be 400 seconds and it does not seem to have plans for steady state operation. Of course, ARC and SPARC expect to do just this. Commonwealth's plans are to drive a steady state current externally with microwaves and rf. However, a great deal of experimental data is now in, and it is not encouraging. The Korean tokamak KSTAR [55] and the Chinese, EAST [56] have both driven current for long periods of time externally. The problem is that it takes a great deal more power to drive the current than would be acceptable for economic fusion. [57,11]

While the final results on external current drive are not necessarily in, and future results may be more encouraging, MIT is currently backing away from external current drive, and is considering having the current oscillate back and forth in direction with a period of perhaps on the order of an hour in each toroidal direction. [58] However, when the current goes through zero, there is no MHD equilibrium, and the plasma will splash onto the walls, virtually instantaneously on the time scale of the current waveform. Who knows how long it will take to clean up the mess? In other words, Commonwealth claims its tokamak will provide power to the grid in in a decade, but there are serious uncertainties about how it will drive the current.

Then there is the issue of heating and (maybe) current drive. It will be driven by ICRH (ion cyclotron resonance heating) in SPARC. Inside the plasma, shown in pink in Fig. (3) in their schematic of SPARC, is the belt used to drive the rf power. However, there is no physical structure between the plasma and



the belt which can protect it from the intense flux of 14 MeV fusion neutrons, which as we have just seen will be nearly a megawatt per m$^2$ if the experiment is successful. Can the belt stand up to such punishment without losing its electrical properties, and perhaps even its mechanical properties? To this author, it seems far-fetched. However, it is not up me to prove that it cannot survive in that environment, it is up to the designers of SPARC to prove that it can. The only way for them to do so is to show such a current carrying wire or belt that has stood up to such a neutron flux for relevant times.

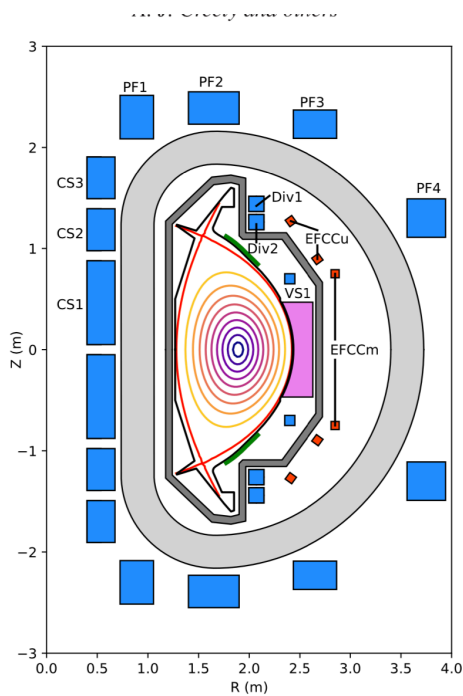

Figure (3): A schematic of the poloidal cross section of SPARC taken from Ref. (50). The pink part inside the vacuum chamber is the belt which carries the current rf current used for ion cyclotron heating.

Seeing the belt inside the vacuum chamber, which must carry a large oscillating current exposed to the neutron flux reminds one of a series of very rough hand drawn cartoons Professor David Rose of MIT circulated in the late 1960's and early 1970's, when I was a graduate student and junior faculty member at MIT.



I do not remember all of them, but the one on rf heating was particularly revealing in this context. Figure (4) is the author's hand drawn recollection of it.

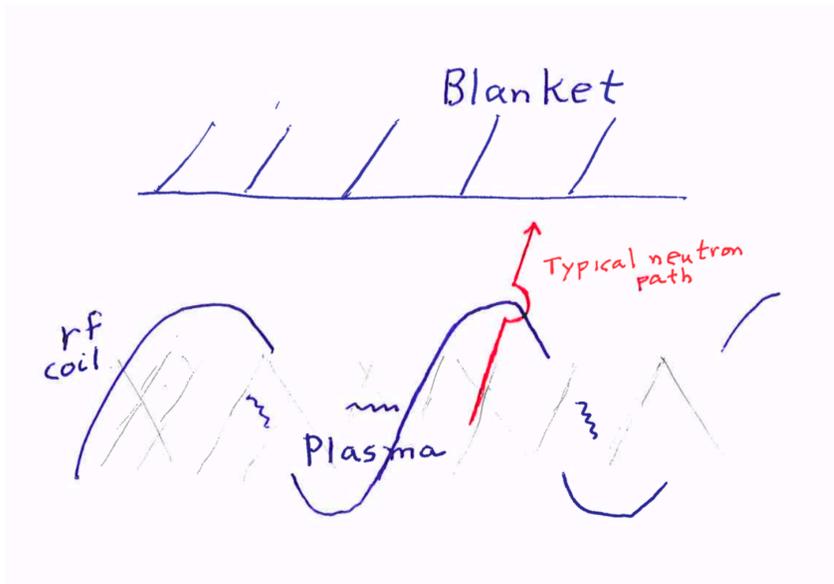

Figure (4): RF heater's view of fusion. Redrawn from memory from a collection of hand drawn cartoons and captions that Professor David Rose of MIT circulated in the 1960's and 1970's.

To this author's mind, the plasma heating (and current drive?) sources should be neutral beams and ECRH. These are the only sources with sufficient standoff. All other sources have key vulnerable components inside the vacuum vessel and are exposed to intense neutron fluxes. Who knows what the effect of a long term $1MW/m^2$ flux of 14 MeV neutrons on these antennas, waveguides, coils or belts will be.

Note also that the thickness of the vacuum wall in SPARC is about half a meter, and this is for a steady state fusion reactor. Conservative design rules (see next section and references specified there ) specify a meter and a half. ITER with lesser wall loading on a pulsed reactor has a thicker vacuum wall. In short, no matter how small you can make the tokamak plasma, the necessity for a wall of ~1.5 meters thick gives a minimum for how small it is reasonable to make the plasma. Neutron leakage out the back may well be a problem for SPARC.



Speaking of heating, if SPARC achieves a Q~10, then 20% of the fusion power will be from 3.5 MeV doubly charged alpha particles which tend to stay in the machine. There has been no experimental work on significant alpha heating in any magnetic confinement device, although undoubtedly there have been paper studies. What is required? Does anyone know? Do all the alphas have to be removed? Some of them? Can their heating be controlled in any way? How would one do this? Surely there should be some experimental work on this issue before barging ahead with an industrial facility reliant on it.

Finally, the design of ARC makes some rather optimistic assumptions of the various efficiencies. It assume an efficiency of 40% for conversion of fusion power to electrical power. While this is reasonable for the most modern coal fired ultra super critical plants, and modern gas fired plants do even better, existing nuclear power plants typically are stuck around a more standard efficiency of ~1/3. Furthermore their rf and microwave sources are assumed to have an efficiency of 42% conversion from source to power injected into the plasma. This is also a rather optimistic assumption. Making a more conservative assumption, and taking the 2 efficiencies of 33% and 25%, their power to the grid drops from ~300MW to ~ 100MW. The design of ARC is walking delicately on the edge of a cliff in parameter space; drop the efficiency numbers a little, and the power to the grid drops a lot.

To summarize, ARC needs 5 miracles in the next 10 years. These are: running steady state with higher wall loading than any other fusion devices including ITER, figuring our how to drive the current, figuring out how to control the alpha heating, operating with vital unshielded components inside the vacuum vessel, and walking right along the edge of a steep cliff in parameter space where a false step could be catastrophic.

Yet as best this author can discern, due to the potential benefit of higher toroidal field, Commonwealth Fusion's plans make make more sense than those of many of the other 'start ups'.

The General Atomic group has also proposed an advanced tokamak based pilot plant, [59] and this author has published an article skeptical of their chances of success, and even suggested that they try breeding instead; it is much easier. [60] The GA response has also been published[61]. To this author's mind, published results like these in the scientific literature is the proper way to adjudicate



important scientific conflicts. I believe that neither I nor GA had resorted to personal attacks.

Another open controversy in the scientific literature involves TAE (formerly Tri alpha energy). At the earliest stages of TAE, this author was involved. Before getting private $$$ to form their company, Rostoker, Binderbauer and Monkhorst attempted to sell it to ONR (office of naval research) as a means of ship propulsion. [62] In their first publication on the subject, they even showed a schematic of a sailor in front of the reactor. Their scheme was a reversed field pinch and their reaction was the p-$^{11}$B reaction. This reaction has much lower cross section than DT, and required much more energetic fuel particles. It produces 3 alpha particles, the reason for the company's original name. They planned to use beams at the energy that maximized the reaction cross section.

ONR turned their proposal to NRL to review, and NRL gave the job to Martin Lampe and me. We spent months studying their proposal, and our conclusion was that their scheme made absolutely no sense. The number of miracles it needed was almost too large to count. We documented our analysis in an NRL Memorandum report [63] but decided not to publish it in the archival literature. The NRL memo is available either from NRL or from the author. ONR decided not to fund the project. It is now 25 years since their article was published, and still no commercial fusion reactor.

Other 'start ups' include many additional plasma confinement schemes. For instance there is a concept of a spherical tokamak. To this author's mind, this has all the problems of the other tokamaks mentioned here, plus a few more. The idea is to shrink a tokamak down to basically a spherical shape. Topologically it is still a torus, but with an aspect ratio of about unity. This means that there will be a thin superconducting center conductor, but carrying the current of all the toroidal coils, and which will be bombarded by the full flux of the fusion neutrons.

Several, including TAE think in terms of a field reversed configuration (FRC). It would be wonderful if this could work, it is an ideal geometrical configuration, but so far their triple fusion product is several orders of magnitude below what JT-60 has achieved [64]. One 'start up' even thinks in terms of this configuration, but with a D-$^3$He plasma, fueled by a D-D reaction, which produces equal amounts of T and $^3$He. But the D-$^3$He reaction has about 10% of the reaction rate of D-T and the D-D another factor of 10 below the .D-$^3$He reaction. It is difficult to see how sufficient D-$^3$He plasma can be generated and maintained so as to dominate the reactions.

Other's think of an FRC, but compressed by an imploding metal liner. This configuration had been analyzed by NRL, and especially Los Alamos over the last 50 years, but has never been built by either lab. Another assures us that we can have heavy ion beam inertial fusion quickly. However, this has been analyzed for over 30 years by the Lawrence Berkeley National lab, but neither the ion accelerator nor the necessary storage rings (enormous components) have ever been built.

In short, the few 'start up' configurations that have been publicly analyzed, are extremely controversial, to say the least. Not one has yet produced even a single 14 MeV neutron, a minimum of $10^{21}$ per second are needed for an economical fusion device. Others have not been analyzed publicly, but seem to need tremendous advances from where we are now.

IV. Pure fusion will not be available this century

Despite the claims of the 'fusion start ups' that they have advantages over the legacy, government supported fusion efforts, these legacy efforts have done pretty well. This section reviews some of the accomplishments of the tokamak and laser, approaches, but also predicts that there are too many scientific, engineering, and economic obstacles for them to develop into power sources hooked up to the grid in this century. Of course, as Yogi Berra said, "Predictions are tough, especially about the future". But consider the following: The century is already 22 years old, and the rate of fusion progress in this first 22% of the century hardly provides encouragement. There are many, many obstacles to get over between where we are now, and economical pure fusion. Obviously many of us would make different predictions, but at this point I am sticking to mine. Fortunately I am old enough so it is unlikely that I will see my prediction fail.

Consider first the tokamaks. After a long break, JET has resumed DT experiments and has produced about double fusion power than in its 1997 campaign, [65] reaching Q~0.4, ~ 10 MW recently. Figure (16) from Ref. (65) shows their earlier and current fusion power.

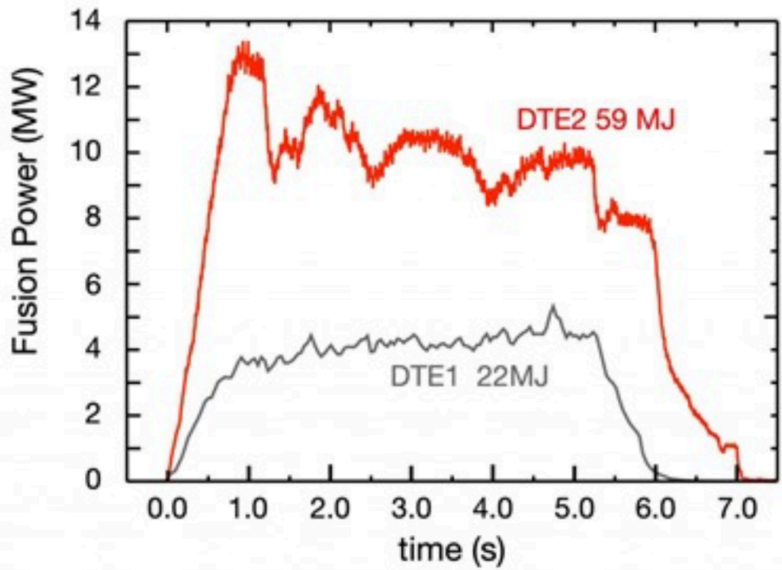

Figure (5): A plot of the fusion power in JET's most recent DT campaign compared to what it accomplished in 1997.

Furthermore, the construction of ITER is now ~ 70% complete. If there are no further delays, it should have its first plasma by 2025, 3 years from now, after a construction period of well over a decade and design period of ~ 20 years. ITER hopes to achieve a Q ~ 10, and generate ~ 500MW of fusion power driven by ~ 50 MW of neutral beams, microwaves, and millimeter waves (56,57).

However, this can hardly be an end point. Electric power is produced with an efficiency of ~ 1/3, so the 500 MW of power means ~170MW sent to the grid. However, beams and microwaves are not produced with 100% efficiency either, once again ¼- 1/3 is a typical number. Hence taking the larger number, the 50 MW of input power would take ~150MW of wall plug power, leaving virtually nothing for the grid.

To make ITER an economical machine, first the gain would have to increase by at least a factor of 3 or 4, so that the circulating power is a much smaller fraction of the total power. Second, the power would have to be increased by a factor of about 5 or 6 to make it comparable to current power stations. Third the size and cost would have to be reduced substantially to make it economically viable. But with larger power and a smaller size, the wall and diverter loading would increase by at least an order of magnitude. These are not minor details! Assuming they could be accomplished at all, they would take decades and additional $10B's. Realistically, pure fusion, at least on the ITER pathway, is at best a 22nd century possibility. However, we now make the case that even this is very unlikely, given current understanding of tokamak physics.

As described in Refs. (11, 32, 35-37) and in their references, and also elaborated in Ref (38), tokamaks are constrained in the current, pressure and density they can contain. Freidberg derived similar constraints.[66] We have called these conservative design rules (CDR's). These are expressed as limits on tokamak parameters:

$\beta_N \leq 2.5$ **(**the normalized beta)

$q_{95} > 3$    (the q at the magnetic surface containing 95% of the plasma current. It is proportional to the reciprocal of the plasma current).   It is also called the safety factor.

$n < 0.75\, n_G$   (the Greenwald density).

The blanket must be at least a meter thick, preferably a meter and a half, to contain all the 14 MeV neutrons and prevent leakage out the back.

The name conservative is used because if they are violated, the result is generally a major disruption.  Disruptions are the sudden loss of energy and/or current.  A tokamak is plagued by what are called minor disruptions, where small amounts of energy escape. However, there are also what are called major disruptions, where virtually the entire plasma suddenly collapses onto the wall.  These have plagued the tokamak program since its birth.  A great deal of effort in the tokamak program has been dedicated to understanding and preventing these.  Generally, this means the tokamak must operate in a region of parameter space defined by these 'conservative design rules'.  This was first pointed out by the author in 2009, [32] and extended by Freidberg et al. [66]
Since a major disruption is obviously something to be avoided in a tokamak reactor, its design has to be conservative as regards these parameters.  These limits are not controversial; they have a solid base in theory and have been confirmed in a wide variety of experiments.  Every large tokamak has been constrained by them.  For instance, the nature of a large number of discharges in JT-60 have been plotted out in the $q_{95}\beta_N$ plane shown in Fig. (6). [36,67]

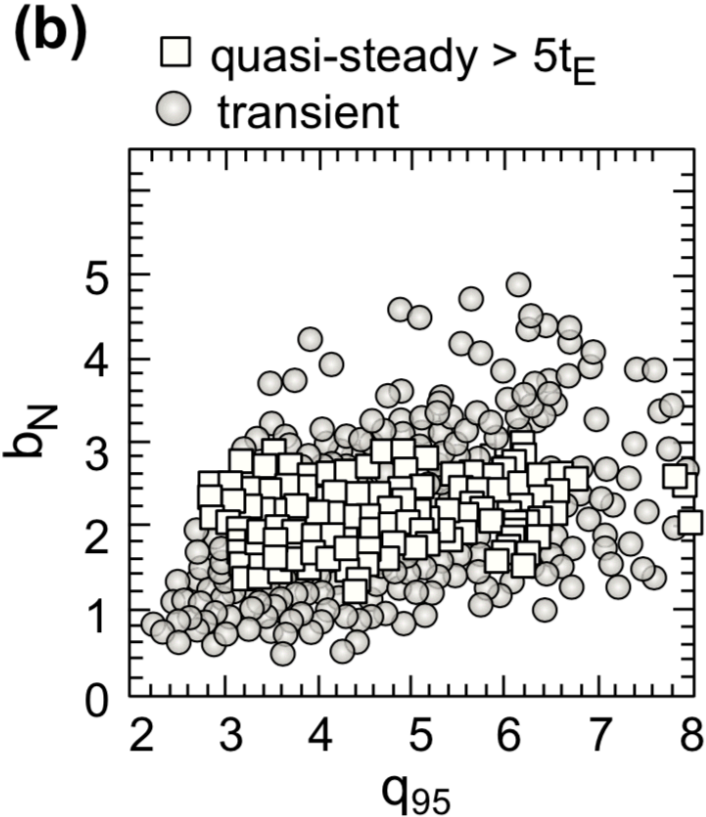

Figure (6) : A plot in ($b_N$ $q_{95}$ space) of aspects of a large number of discharges in JT-60. The blank squares represent discharges that have lasted at least 5 energy confinement times, the shaded circles are discharged that did not, i.e. discharges marked as transient. Clearly steady state is only achievable for $q_{95}$ > 3 and $\beta_N$ < 2.5, just as predicted by conservative design rules.

Notice that JT-60 has also gotten discharges with high $\beta_N$, as high as 5 for $q_{95}$ of 6 and for $q_{95}$ as low as 2, for $\beta_N$ about 1. However, these constitute no improvement in the actual beta, which is proportional to $\beta_N$ times the current, or to $\beta_N/q_{95}$. In addition, these all disrupt in less than 5 energy confinement times. The European tokamak JET also showed very similar results. [68]

This author has derived two expressions for the fusion power based on the maximum that CDR's would allow. [38] They depend on assumptions of profiles and temperature ratio, and are

$$P < 0.06 K^2 [aB]^4/R \qquad (1)$$

$$P < 2 \times 10^{-3} \pi^2 KRa^2 B^4 \qquad (2)$$

Now let us consider a 3GW tokamak (like a conventional power plant), taking K=1.7 (the ellipticity in the poloidal plane), and a = R/3. Then Eqs. (1-2) become formulas for the minimum radius R, as a function of the magnetic field. These are summarized in Table (II). The left-hand column specifies the magnetic field, 5 or 9 T, the top row, R, from Eqs. (1-2). The Table entries are the minimum radii in meters (rounded to the nearest whole number).

| B | R [Eq(2)] | R [Eq(1)] |
|---|---|---|
| 5 | 11 | 12 |
| 9 | 5 | 6 |

TABLE II: Minimum major radius for a 3 GW tokamak as calculated by CDR's for fields of 5 and 9T.

Consider a 5T power plant. The entire device, counting the field coils and shielding would extend from the goal line to about the 30–35-yard line of an American football field. It simply does not seem that this would be affordable for every power station. A fast neutron reactor seems like a much more affordable choice.

Furthermore, there are a considerable number of other vital issues, for which ITER, at this point, has no answers and apparently no specific plans to address these issues experimentally. These are how to drive the current steady state or at high duty cycle, and finally how to make the gain larger than 10 (the ITER gain) by at least a factor of 3 or 4, so that the circulating power is not such a dominant fraction of the total power, i.e., so significant power is left over for the grid.

Given the size and scale of such a 3GWth (1GWe) tokamak, as well as considering the time it has taken to design and build ITER (~40 years), as well as the cost and scientific uncertainties involved, it does not seem reasonable that such a machine can be developed, economically put in place, and become a dominant source of electrical power by century's end.

Furthermore, ITER has hit some new delays [69] and the time for first plasma is uncertain, but has certainly slipped. Among other things, large components, manufactured in different countries, with different leadership and different manufacturing cultures (i.e., like herding cats), have to fit together with micron tolerance, and in ITER, they do not.

"The vacuum vessel sections are 17 meters tall and 6-7 meters wide. 'You have to fit them together to a fraction of a millimeter' Befouled says". [69]

While this is certainly discouraging, it has a silver lining. An ITER like commercial fusion breeder or fusion device will certainly be manufactured under the auspices of a single national or commercial leadership. Components will most likely be much better coordinated, and this alone will almost certainly reduce mismatches as well as the costs.

Given the results of the recent LLNL (Lawrence Livermore National Laboratory) experiment, [70] there is a new kid on the block, laser fusion. For years, except for Ref. (34), this author did not include laser fusion in his work on fusion breeding, as it had failed to produce anything like the neutrons necessary to be considered a potential energy source. Furthermore, full disclosure, the author is a participant, as a consultant, in the NRL laser fusion program. In fact, laser fusion was never supported for fusion energy research, but for nuclear stockpile stewardship. With the result of the August 2021 experiment, that has all changed, and laser fusion must now be regarded as a serious contender. For the last few years, the LLNL project has been inching up on what they define as a burning plasma, with this or that definition for their goal. As they advanced, they got Q's nearly as high as 10%. However, while they might have defined the plasma as burning, this depended on subtle interpretations of their measurement. Typically, they measured the neutron energy emitted, surmised the alpha particle energy produced (which was absorbed locally), and then calculated the PdV energy going into the fuel to heat it, and if it was less than the alpha energy, they claimed a burning plasma. Even with this only 'sort of convincing' justification for their claims, their paper describing these measurements (70) was downloaded over 80 *thousand* times!

All that changed with their August 2021 shot. To their surprise (some of their diagnostics were set for lower fluxes and saturated!) and delight, they achieved 1.3 MJ of fusion energy from 1.7 MJ of laser light energy. [71] Not only that, they achieved a non-subtle signature of a burning plasma. They measured the time dependent temperature of the exploding fuel and ablator, and saw that as it expanded, it *heated*. There is no explanation for this result but the fact that as the plasma expanded, it was heated by the alphas and a burn wave was set up. It was most definitely an ignited plasma, nearly 20 years before ITER hopes to achieve ignition. To this author's mind, it was a Wright Brothers moment. Unfortunately, these results had not been published in the archival literature, in time for this author to cite them, but unquestionably they soon will be. However, the LLNL group gave a series of several zoom seminars on their results, which many, including the author watched (the author has not been able to recover these despite several Google searches).

Now that laser fusion is on the scoreboard, let's add up the score. While the results are nothing if not amazing and impressive, economic laser fusion power is hardly around the corner. Even with the Wright Brothers achievement, it took another half century to develop large jet powered aircraft. For one thing, while the Q is 0.67, greater than the JET Q of 0.4, a more reasonable measurement for energy production is the efficiency of the driver $\eta$, times Q. For JET, $\eta Q \sim 0.1$, while for $\eta$ for the NIF is probably around 1%, so $\eta Q \sim 0.0067$.

The target in the LLNL experiment is inside a container called a hohlraum. Their laser is not focused on the target, but on the inner hohlraum walls, which emit X-rays which impinge on the target and implode it. As the goal of the project is nuclear stockpile stewardship, and not energy, the sponsor is only interested in X-ray drive and is not interested in things like efficiency, rep rate, or bandwidth, parameters important for energy. As the laser light does not directly strike the target, this configuration is called indirect drive. Figure (7), taken from Ref. (71) is a schematic of LLNL's configuration.

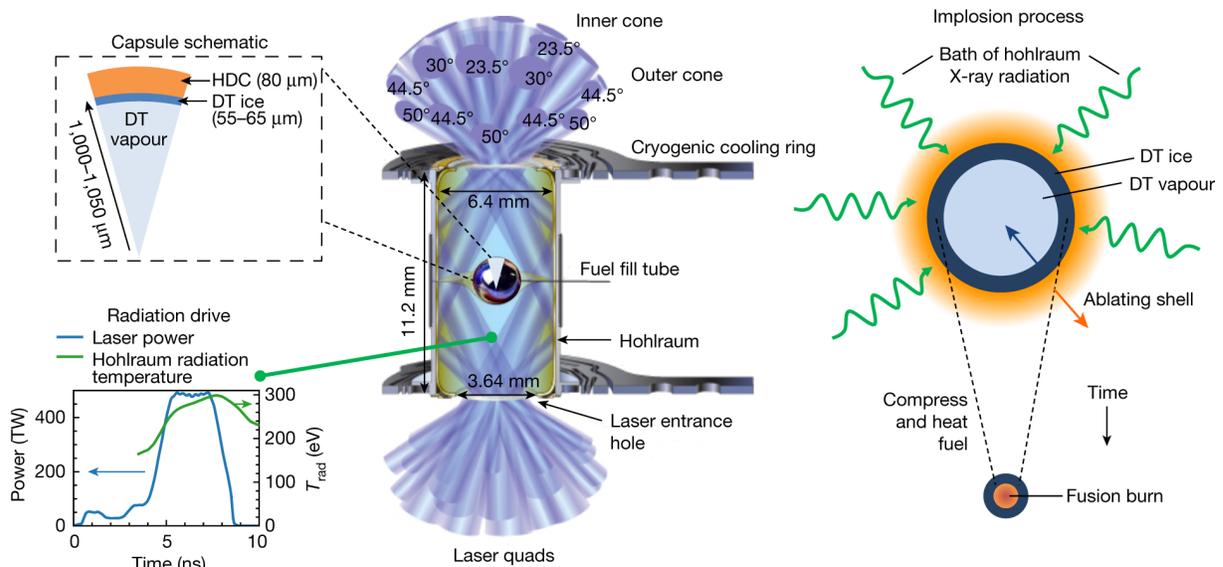

Figure (7): A schematic of the LLNL approach to laser fusion taken from Ref (71).

The path LLNL is on has quite a few problems if the goal is energy rather than nuclear stockpile stewardship. First of all, each shot involves a hohlraum, a precisely engineered container, made with expensive materials like gold or uranium and currently costing thousands of dollars each. While mass manufacturing of hohlraums will undoubtedly bring down their price considerably, even if the target produces a total energy of ~100MJ, which would translate to ~33MJ of electric energy, or ~ 10 kWhrs, worth about a dollar, it gives a very low-price limit for the ultimate economically acceptable hohlraum price. Second of all, only a small fraction of the laser light (in the form of X-rays) makes it to the target; the rest is lost through other channels. This is shown in Fig (8) taken from the LLNL publication (72).

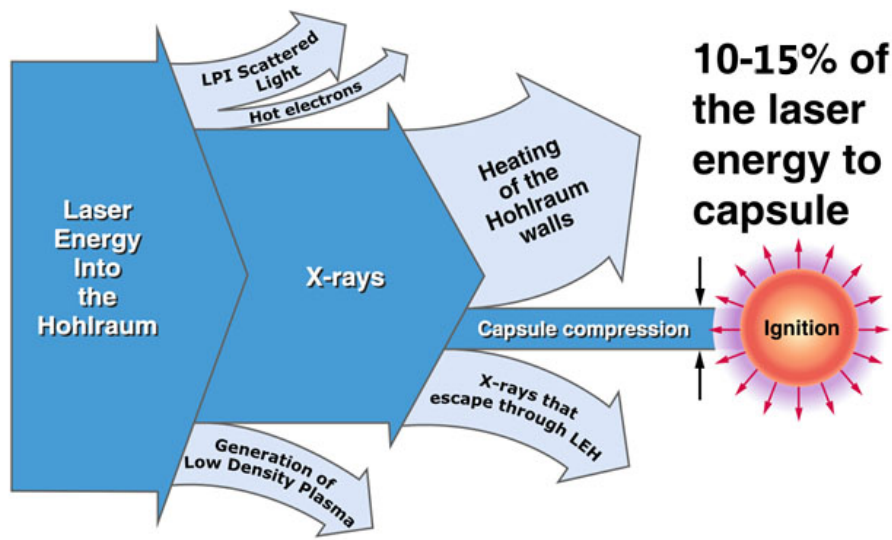

Figure (8): A schematic of where the laser energy goes for an indirect drive configuration. Only 10-15% of the laser energy makes it to the target in the form of X-rays.

Finally, the LLNL configuration is fine for one shot, with the target on a small stalk, and focusing the laser on it is relatively simple. It is rather like hitting a golf ball on a tee. To do this continually, targets would have to be continuously shot in a high speed, with each shot certainly traveling in on a slightly different path. The target engagement becomes more like hitting a variety of Jacob deGrom's fastballs, curve balls, sliders, changeups….., *on every pitch*. Not only does the target have to be in the right place, it has to have the proper orientation also, so the laser is aligned with the axis of the hohlraum, or to use the baseball analogy a bit further, the batter has to hit the ball at a precise phase of the ball's spin.

An alternative approach is what is called direct drive, where the laser light directly hits the target, and without any of the losses specified in Fig (8). Figure (9) is a schematic of direct drive configuration taken from NRL. [73]
Since the target is a sphere, it does not have to have any specific orientation, so the target engagement becomes much simpler. LLNL is not set up for uniform $4\pi$ illumination, but the University of Rochester Laboratory for Laser Energetic (URLLE) has done direct drive experiments with its smaller OMEGA ($\Omega$) laser, [74] and NRL has done a good bit of theory on it. [75-77]

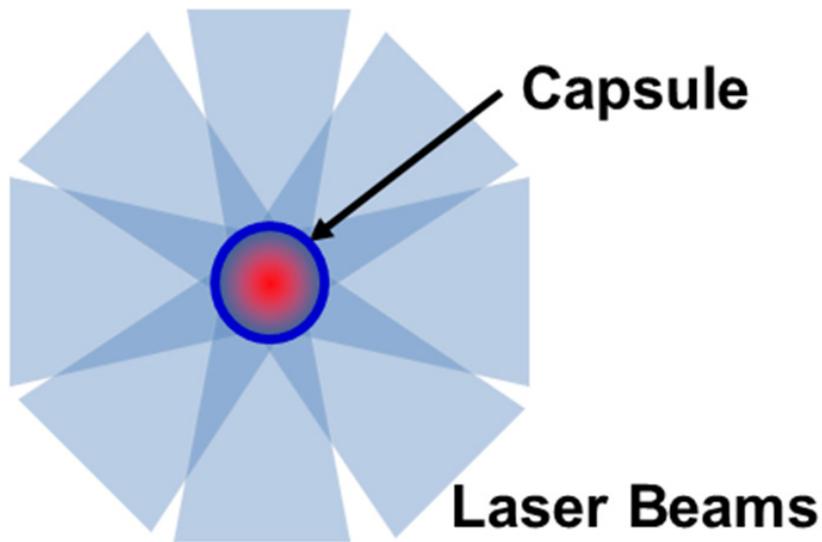

Figure (9): A schematic of direct drive laser fusion taken from Ref. (73). The laser beams directly hit the target, so very little energy is wasted in other loss channels, as is the case with indirect drive.

Since so much less laser light is wasted in direct drive, calculated gains (i.e.~ hundreds) are generally considerably higher than those gains calculated with indirect drive (i.e ~10) . Figure (10) from NRL [75-77] shows a variety of gains from a variety of target configurations and laser pulse structures.

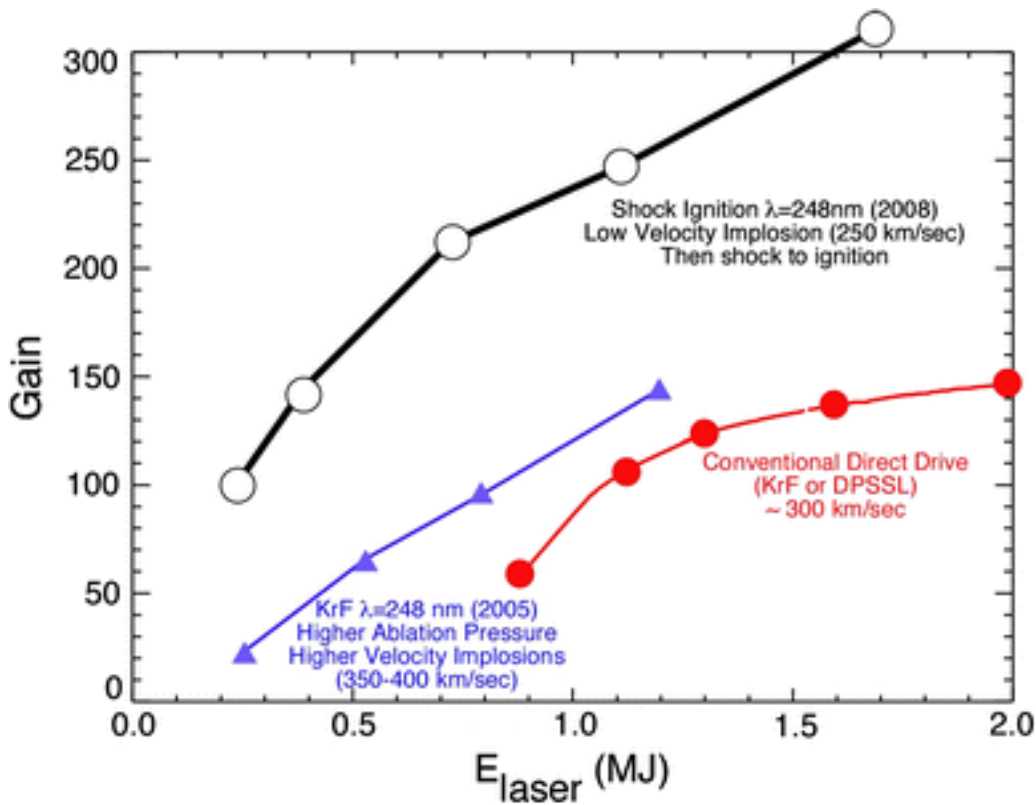

Figure (10): A variety of calculations of gain from NRL for direct drive laser fusion.

Let us then set up two straw men for laser fusion. First say that one develops a 5% efficient, 1 MJ laser, and has a target with a gain of 100. This then gives 100MJ of fusion energy, translating to 33 MJ of electrical wall plug energy. However the laser will gobble up 20 of those MJ's just to run itself, so this would hardly be economical.

The second assumes a target gain of 200 and a laser efficiency of 10%. Here a 1 MJ laser would give 200 MJ or fusion energy per pulse, or 66 MJ of electric energy, and when 10 are subtracted to power the laser, 56 MJ of electric energy per pulse delivered to the grid. This would most likely be a viable power source. The device would have to operate at ~ 20 Hz to generate 1 GWe as a normal electric power plant.

But what does this mean for the development of economic laser fusion power this century, 78 years from this writing? First of all, one needs much more efficient high power lasers with greater bandwidth and rep rate.  Then one has to illuminate a target and show that the gain calculation such as those in Fig (10) really do make sense.  So far the best calculations have fallen far short in predicting gains (as pointed out, earlier LLNL estimates were 10, but it took them a decade to get to 0.67!)

Many of these issues were studied by a project, run by NRL, called HAPL [78] (High average power laser) which lasted about 10 years until it was cancelled in 2008.  It was a multi institutional project which examined many such obstacles, including target manufacture, tracking, and engagement, final optics, laser development, target chamber, tritium production, recovering unburned tritium….

In the years of its existence, it made very good progress on many of the issues.  For instance both NRL and LLNL developed preliminary versions of rep rated lasers which they saw as possible to develop into lasers relevant for laser fusion.  Of course, given its short life time, HAPL had not completed the job.  However it found no show stoppers.   There does not seem to be anything like tokamak conservative design rules, which greatly constrain what  tokamaks can and cannot do, and pushes these reactors way up in size and cost.

Assuming everything works as well as the HAPL project hoped and expected, can it all be accomplished, so that laser fusion power plants can be installed and become operational on a large scale before the end of the century?   This author feels the answer is no.   The job just seem too big and difficult.  NIF took about half a decade longer to construct than was initially planned.  The national ignition campaign (NIC) was to end in 2012 with a Q of ~10.  Instead it took an additional decade to get a Q of 0.67.  Already  past delays have gobbled up one and a half of the not quite 8 decades available and the project  still missed its original goals by more than an order of magnitude.  Furthermore LLNL's next few shots were unable to repeat the high gain.  It took them over a year to get 1.2 MJ of fusion products with a 1.9 MJ laser energy.  Judging by the history up to now, this author's opinion is that there are just too many orders of magnitude to go and too little time to get there, to reach economic power production and install it on a large scale this century.

## V. Fusion Breeding for mid century sustainable worldwide power

We first motivate the author's claim that fusion is the 'best breeder'. It is really very simple.

The fission reaction directly produces 2-3 neutrons with an energy of ~ 2MeV. For fission breeding, one of these neutrons is needed to continue the chain reaction and one is needed to replace the burned fuel atom. Since there are losses, at most half a neutron per reaction is available to breed. In other words, at maximum breeding rate, it would take two breeders to fuel one thermal reactor of equal power.

Now consider fusion. As we will see shortly, the after breeding the tritium to fuel fusion reactor, there are still sufficient neutrons for breeding $^{233}$U, probably about ½ to 1 from each fusion reaction. However, the fusion produced neutron energy is 14 MeV, while the energy released from the $^{233}$U fission reaction it fuels is ~ 200 MeV. Hence a single fusion breeder could fuel 5-10 thermal reactors of equal power, making it far and away, 'the best breeder'.

Let us now consider the breeding of tritium. The conventional breeding reaction is usually assumed to be

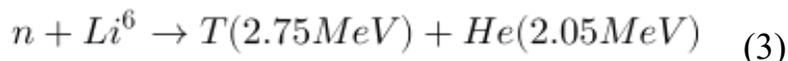

$$n + Li^6 \to T(2.75 MeV) + He(2.05 MeV) \quad (3)$$

This has the advantage of being exothermic, adding energy to the reaction, but the reaction costs a neutron. However if there are any neutron losses at all, an inevitability, the reaction will ultimately peter out. Even for pure fusion, a way must be found to produce additional neutrons. For fusion breeding, where more neutrons are required, the need is especially acute.

First of all, the fusion produces a much more energetic neutron, and an energetic neutron can produce more neutrons in particular targets. At an energy of 14 MeV, the neutron can produce several others, as is shown in Figure (11) for a lead target. [11,79]

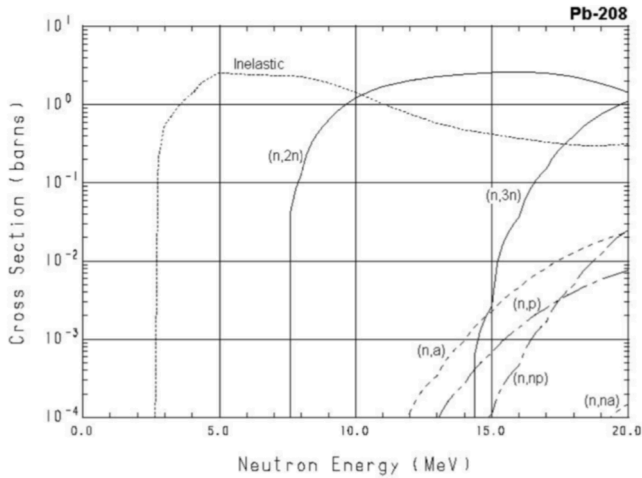

Figure (11). The cross sections for producing 1 or 2 additional neutrons in a lead target, as a function of the incident neutron energy.

An even more prolific neutron source is beryllium, which requires a neutron energy of only 2.7 MeV to produce another neutron. Not only that, there is an additional tritium breeding reaction which conserves neutrons. It is

$$n + Li^7 \to T + He + n - 2.47 MeV \qquad (4)$$

This reaction is endothermic, and costs about 4.5 MeV as compared to the reaction in Eq. (3). However compared to the extra neutron, this energy loss is of no importance if the goal is fusion breeding. Let us say that after losses, the extra neutron breeds an extra half $^{233}U$ nucleus. When burned it produces ~ 100MeV, much more than making up for the extra energy needed to produce the extra neutron. Hence there are many possibilities for producing sufficient neutrons for breeding $^{233}U$ as well as tritium from the initial fusion reaction.

Now let us consider the breeding process for $^{233}U$. The collision cross section for both fission and neutron absorption for $^{235}U$ and $^{238}U$ is shown in Fig.(2).

In general actinides with odd atomic mass have much larger fission cross sections at thermal energy, even atomic mass actinides do not. A fission breeder must work with the 2 MeV neutrons, while a thermal reactor burns neutrons with much greater reaction cross section. This renders a fast neutron reactor much more complicated, expensive, and one which requires much more fissile material to get started. [43]

Now let us look at thorium - neutron reactions. When a neutron collides with a thorium nucleus, the absorption reaction has a cross section very much like uranium, shown in Fig. (2). It goes from $^{232}$Th to $^{233}$Th. The new new nucleus is unstable, with a half life of about 20 minutes, to a single beta decay, where it becomes Protactinium, $^{233}$Pa, which is itself unstable with a half life of about a month. $^{233}$Pa beta decays and become $^{233}$U, which is stable. However $^{233}$U, having an odd number atomic weight has large thermal neutron fission cross section, very much like that shown in Fig. (2) for $^{235}$U. In other words it is fine as a fuel for thermal nuclear reactor. There is a similar reaction chain starting with $^{238}$U and ending with $^{239}$Pu, but since we would like to avoid plutonium to the extent possible, we do not consider this. Figure (12) is a schematic of the fusion based $^{233}$U breeding process.

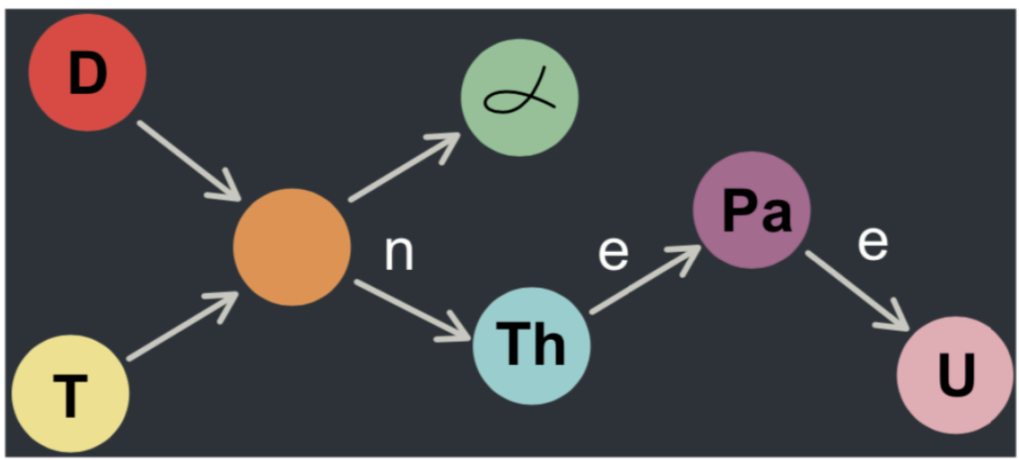

Figure 12: A schematic of the decay process where a fusion neutron is absorbed by a thorium nucleus, setting into motion a decay process which finally ends up as a $^{233}$U nucleus; a perfectly good fuel for a thermal nuclear

reactor.

It is important to note also that the reactions which take one from $^{232}$Th to $^{233}$U are exothermic, and roughly double the neutron power of the fusion reaction. To analyze the fate of a 14 MeV neutron entering a target, one uses Monte Carlo codes which all the main DoE labs have. Table 14.5 of Ref. (18) gives some examples of the energy released and the particles produced from a 14 MeV neutron entering particular homogeneous materials. A portion of that table is reduced as Table (III).

| Medium | Product atoms | Energy released (Mev) |
|---|---|---|
| $^{232}$Th + 16% $^6$Li | 1.3 $^{233}$U + 1.1 T | 49 |
| $^9$Be + 5% $^6$Li | 2.7 T | 22 |
| $^9$Be + 5% $^{232}$Th | 2.66 $^{233}$U | 30 |
| $^7$Li + 0.8% $^{232}$Th + .02% $^6$Li | 0.8 $^{233}$U + 1.1T | 17 |

TABLE (III) Product atoms and energy released by a 14 MeV neutron impinging on various homogeneous materials

Hence ITER, which is designed to produce ~ 500MW of fusion power, would, as a breeder, produce ~ 1GW; and the original Large ITER [80], designed to produce ~ 1.5GW, would, as a breeder produce ~ 3GW, about equivalent to a modern coal or nuclear powered electric generating plant. Many more details of the reactions as well as possible reactor designs are provided in Refs. (81-83). One design, using a realistic blanket geometry [82], would produce about 0.6 $^{233}$U atoms, as well as the necessary tritium atom, from every 14 MeV neutron. However when burned, this produces ~120 MeV, multiplying the neutron energy by about an order of magnitude. Hence one breeder can fuel about 10 thermal reactors of equal neutron power, or about 5 of equal total power (recall the breeding reactions are exothermal and roughly double the the reactor total power).

Now let us do a very rough estimate of the cost of the fuel produced. This is based to a large degree on what the cost of an ITER scale reactor would be. Unfortunately, the cost of ITER has been increasing very rapidly, and not only is this discouraging, it makes an estimate difficult. The original cost of Large ITER was to be $10B in capital cost and $10 in operating cost for 10 years. Let us assume that the capital cost of the commercial prototype is $25B. The machine is assumed to last 30 years. Let us assume the same billion dollars per year operating cost.

Thus, as a very rough estimate, let us say the capital and operating cost of the commercial prototype is $2–2.5B/year. It is a reactor, which generates 1GWe. Assuming it runs all year, and sells the power for ten cents per kWh, it earns about $0.9B. But it also produces 5 GWe of nuclear fuel. To recover the additional $1.1B, it would have to sell the nuclear fuel for about 2–3 cents per kWhr. This estimate is certainly not exact, and as costs capital and operating costs of ITER become clearer, it can be revised.

But at this point, the estimated cost does not seem to be any kind of show stopper. Uranium fuel for LWR's now costs about one half to one cent per kWh, so fusion bred fuel might increase the electricity cost by a penny or two per kWh.

To summarize, there does seem to be a roadmap to large scale, economic power production via magnetic fusion breeding by mid-century. Pure fusion can claim no such magnetic roadmap at this point. Pure inertial fusion might, as there are no conservative design rules that we know of holding it back. However, IFE still has to get over significant hurdles to get the neutron production that MFE has right now. Unquestionably, fusion breeding is a more conservative goal for IFE than is pure fusion; perhaps it is the only reasonable goal.

One aspect of fusion breeding is worth mentioning is that the fusion device would need a flowing liquid blanket. One possibility for the liquid is a molten salt such as FLiBe, which has lithium to produce the tritium and beryllium as a neutron multiplier. Thorium, protactinium and uranium are all soluble in it.

The liquid flows in and out, and when it is in the machine, the neutrons can react to produce the Pa and T, and these can be extracted chemically as the liquid flows out of the machine. One schematic of such a reactor is shown in Fig. (13) from (83).

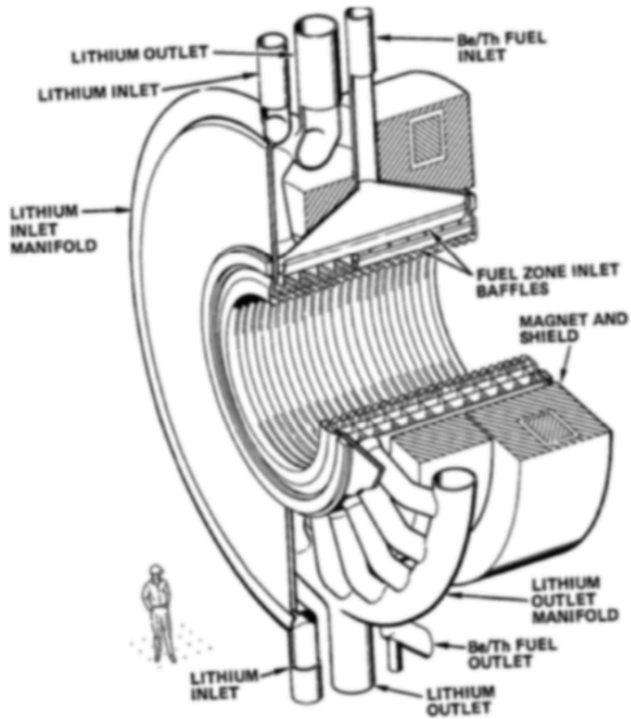

Figure 13: A schematic of a fusion breeding blanket surrounding a fusion reactor. Notice the input and exit pipes for the flowing lithium and thorium.

VI.  Preliminary development plans for achieving fusion breeding by mid century

In this section, we examine the potential routes to achieve economic fusion breeding not too long after midcentury. Unfortunately, the fusion community has confidence that they have a 'perfect' energy source for future civilization.

They would prefer not to tie their fortunes to fission, which might not even want them, and which they see as having issues of safety, proliferation, and long-term radioactive waste. But they should consider realities. Fusion breeding is at least an order of magnitude easier to pull off than pure fusion power. It should be possible, whereas commercial application of pure fusion may turn out not to be. Even in the best of circumstances, commercial fusion breeding should be available quite a few decades earlier than commercial pure fusion.

Hence right now, the main obstacle for the fusion community to get over is a psychological one. It should realize that fusion breeding is at least as good a goal, and possibly even a better one, than pure fusion. It is certainly one that it more achievable, likely not too long after midcentury, and by then, especially as the less developed parts of the world develop, it may be desperately needed. Also, while there is a real need for mid and late century energy sources, there is no need to panic. If fusion breeding can come on line in 30-50 years, that would be fine. Furthermore, it is a goal that fits in much better with current, and almost certainly the future, nuclear infrastructure. After all, it only has to produce fuel for existing and future reactors. It does not have to set up an entirely new energy architecture, which historically has taken quite a few decades.

The author, over the years has been in contact with several nuclear experts, including George Stanford (deceased 2013), one of the principal developers of the Integral Fast Reactor (IFR). Here is an excerpt from an email he sent me [84]:

Fissile material will be at a premium in 4 or 5 decades.....I think the role for fusion is the one you propose, namely as a breeder of fissile material if the time comes when the maximum IFR breeding rate is insufficient to meet demand.

Another nuclear expert was Dan Meneley (deceased 2018), in charge of the Canadian nuclear program, and worked on both the CANDU reactor and the IFR. Here is a comment he sent me upon receiving all of the viewgraphs for a meeting he was running : [85]

We (I'm on the Executive of the Environmental Sciences Division of the ANS) held a "Sustainable Nuclear" double session at the ANS Annual in Reno a couple of weeks ago. I have copies of all the presentations. ............ The result was an interesting mixture of "we have lots", just put the price up and we'll deliver (we've heard the same from Saudi recently) and "better be sure you have a long-term fuel supply contract before you build a new thermal reactor".

and

I've nearly finished prepping my talk for the CNS on June 13th (2006) -- from what I can see now, we will need A LOT of <u>fissile isotopes</u> if we want to fill in the petroleum-energy deficit that is coming upon us. Breeders cannot do it -- your competition will be enrichment of expensive uranium, electro-breeding.

Perhaps George Stanford and Dan Meneley were right. Perhaps the world will need a great deal more nuclear fuel than is available to continue to spread the widespread benefits of civilization. The top down approach to energy supply (Section II) certainly would indicate this. Not too long after midcentury, our descendants might be thanking us if we begin to develop fusion breading NOW.

The rest of this section is divided into three parts. A. First it argues that laser fusion should be treated on (at least) an equal par with tokamak fusion. It presents a variety of arguments for this.

Then B, a possible development route for tokamaks, and C, a possible development route for lasers.

    A. General discussion of inertial and magnetic fusion:

Here we argue for DoE in the USA to treat laser fusion on, at least an equal level, with tokamak fusion. This would mean, at the very least supporting several additional laser development programs and the related experimental work. Just as in the US there used to be several tokamaks, i.e., at PPPL, ORNL, GA, MIT, Texas and UCLA, and now there is only one at GA, and it is not the largest; there should now be several American laser development programs. If we do not build tokamaks, the rest of the world will, but at least right now, we are the only ones with the potential to do laser fusion. Here we present some of the advantages we see for laser fusion as opposed to tokamak fusion.

1. Perhaps most worrisome, tokamaks still do not know how to drive the current, whether in a steady state or pulsed mode. The experiments for external current drive from EAST [56,57] and KSTAR [55] can drive the current externally, it just takes much too much power [11,57]. Hopefully EAST, KSTAR and other superconducting tokamaks will find a way to externally drive the current at lower power  The alternative seems to be an alternating current, but when the current passes through zero, there is no MHD equilibrium, and the plasma will immediately hit the wall. References (11) discusses this.

2. Second, laser fusion is much safer. ITER will have a poloidal field and plasma, which we do not understand very well, and is subject to disruptions, with a stored energy comparable to a ~100-pound bomb. The superconducting magnet will have the energy of a ~1000-pound bomb. An uncontrolled quench, which happened at CERN a few years ago, would cause enormous damage. In CERN, it was in a miles long tunnel, so the damage was mostly local, but still took a year to repair. In the confined space of ITER, an uncontrolled quench would probably bring down the building and much more. Laser fusion does not store this kind of concentrated energy anywhere. For instance, NIF has 192 lasers, each of which has its own pulse power supply supplying perhaps a Megajoule. If one blows up, the damage will be localized and probably would not spread to others. A more efficient laser for a power plant would have even less stored energy.

3. MFE has to worry about confining alphas, laser fusion does not. Experimentally, MFE is nowhere on this and probably won't be for more than a decade or two. Undoubtedly there are lots of paper studies on it. Laser fusion does not regard alphas as an afterthought. Alphas are built into the laser fusion culture at the ground level. Its entire rationale is to set up a burn wave, which NIF has already produced. MFE is nowhere near getting such a result. And once ITER does achieve a burning plasma, how will the heating be controlled? Do the alphas have to be removed? All of them? Some of them? Laser fusion does not have to tackle these very complicated issues.

4. Tokamaks, and any magnetic confinement scheme will always have to worry about recycling as 14 MeV neutrons, radiation, as well as fast ions and neutrals hit the wall and diverter plates. Who knows what is entering the plasma? In laser fusion, whatever hits the wall and bounces back, by the time it gets to the target, the fusion reaction will be long over.

5. Tokamaks are constrained by conservative design rules (CDR's), meaning that any tokamak pure fusion device will be quite large. Most likely, in the light of CDR's, there is no way to shrink the size of the tokamak and still have it give enough power for pure fusion, and even a tokamak breeder will be quite large. At least as far as we are aware now, there are no such theoretical constraints on laser fusion. In fact an ICF burn wave has now been demonstrated at both the mega Joule and megaton level.

6. Laser fusion has flexibility in where it places the chamber wall, tokamaks do not. If the the wall loading is too much, in laser fusion, it can be moved back a bit. Of course moving the wall back, makes the target tracking and engagement more difficult. On the other hand, if the problem is that tracking and engaging the target is too difficult, the wall can be moved in a bit.

7. Laser fusion hardly has a clear glide path to commercial power development. It has to worry about laser efficiency, rep rate, bandwidth, cost, target engagement and manufacture, final optics, target stability, ….. These were all discussed in (77). To this author these various obstacles to laser fusion seem to be more technical in nature, whereas the obstacles tokamaks discussed above, seem to be more fundamental in nature. Tokamaks still need a huge international consortium, costing tens of billions to do in 15 more years (i.e. create a burning plasma), something which LLNL has already done for a small fraction of the cost of ITER.

    B.  A plan to bring tokamak fusion breeding on line by ~ mid century

There are two immediate things the tokamak program should embrace. First it should realize that if ITER is successful, large ITER probably would have been also. Counting the breeding reactions, it could be a 3GWth, 1GWe breeder which could fuel 5-10 thermal nuclear reactors of equal power, as discussed in the last section. ITER becomes an end in itself, not a means, on a long path to who knows what DEMO, to be put in place who knows how many decades later and for who knows how many additional $10's of billions, assuming it can be done at all.

Second, ITER should realize that it seems to have a serious problem with current generation.  Possibly external current generation will ultimately be successful, but right now it takes too much power.  A backup plan seems to be required, and probably the only reasonable backup plan is the use of an alternating Ohmically driven current.  In any case, this is a serious problem requiring ITER's full attention.

Something that ITER plans is to have are 6 ports, which 6 of the partners can use to explore tritium breeding. [86]  This author certainly recommends that at least one of these partners (preferably all in fact) use the $^7$Li pathway to breed the tritium.  This way they preserver the neutron for other applications, especially for breeding fissile fuel.  Perhaps one or more of the partners could also use their particular port to produce some $^{233}$U on a small scale.  This way the fusion project would produce something, admittedly on a very small scale, that the world could actually use.

Hence ITER should continue pretty much on the path it is on, with consideration given to the two points above.  It should not expend any further resources on the DEMO, which probably will never become an economical power supply and almost certainly will never be built.   Instead is should dust off its plans for Large ITER, but alter them so that it could be used with a flowing blanket, probably a molten salt.   Then it would be ready for construction assuming ITER is successful.  Likely, if ITER could also solve the current problem, the world would be ready to seriously begin implimenting fusion breeding on a large scale.

There is also a great deal that can be accomplished with smaller tokamak experiments.  For decades this author has argued that the tokamak program in the United States should build a tokamak he called 'The Scientific Prototype'. [29,87]  This is a tokamak about the size of TFTR, JET or JT-60, 2 of which have already produced large numbers of 14 MeV neutrons.  The idea of the scientific prototype was simply to reproduce this in steady state, or at high duty factor, in a DT plasma,  and also to breed its own tritium and to recover its unburned tritium for reinsertion into the plasma. [11]  If the scientific prototype were built, and was successful, it would have given crucial high duty cycle data; ITER, if successful, would  give Q~10 data.  If both were successful, the  world could embark on a tokamak based fusion breeder program, decades before pure fusion could be ready, assuming it could ever be ready at all.

Unfortunately the DoE did not adopt this path, and neither did PPPL. Had PPPL instead adopted the scientific prototype between say 2000 and 2010, by now surely have confronted the current drive problem, and perhaps would have solve it one way of another. By the time ITER had completed its job on a pulsed machine, a tremendous amount would have already been learned on steady state or high duty cycle operaton. In a worst case scenario, it could have alerted ITER to the fact that steady state or high duty cycle operation of a tokamak had serious unresolved issues. Perhaps it is still not too late to build the scientific prototype, if not in the US, perhaps somewhere else in the world.

Most of this author's advocacy of the scientific prototype assumed external current drive. However on becoming aware of the discouraging EAST [56,57] and KSTAR result, [58] his most recent efforts [11] have explored an oscillating Ohmically driven current. The goal was to start with a duty cycle of ~20% and work up to ~80%. Perhaps there is something to this approach.

At this point, unlike France, China, and Korea, the US has no superconducting tokamak. However it looks like we will soon have one by Commonwealth Fusion, one that can operate at any magnetic field between 3 and 12T. Perhaps once its tokamak becomes operational, Commonwealth Fusion should shift its priorities to solving the steady state current drive problem, either by external drive at an acceptable power, or by an oscillating Ohmically driven current, or by some other means they devise. They will not be giving their investors the pilot plant they promised, but they will not be giving them this pilot plant, in a decade, no matter what they do. Rather than spending the next decade tilting windmills, perhaps they should use their expertise to solve a real and immediate problem in tokamak fusion, namely finding an acceptable way to drive the current. If successful, this would be a genuine and vital contribution they could make in the next decade. They seem to have the equipment and expertise to tackle it as well as anyone else, at least anyone else in the US.

C. A plan to bring laser fusion breeding on line by ~ mid century

While this paper is optimistic as regards laser fusion, especially for breeding, there is little experimental evidence, or plans, to suggest which way to go. In Section V, it was shown that while a 10% efficient laser with target gain of 200 would be fine for pure fusion, a 5% efficient laser with a target gain of 100 would not.

What is vital is to set up laser and target programs to see what is feasible and what is not. This author here proposes a three pronged attack on laser and target development. Two of these prongs were proposed by Steven Bodner in a letter to the National Academy of Science, [88] a letter that should have received much more attention than it did. Figure (14) shows a schematic, taken from Ref.(88), showing the plan regarding both laser development, as well as the target and chamber design.

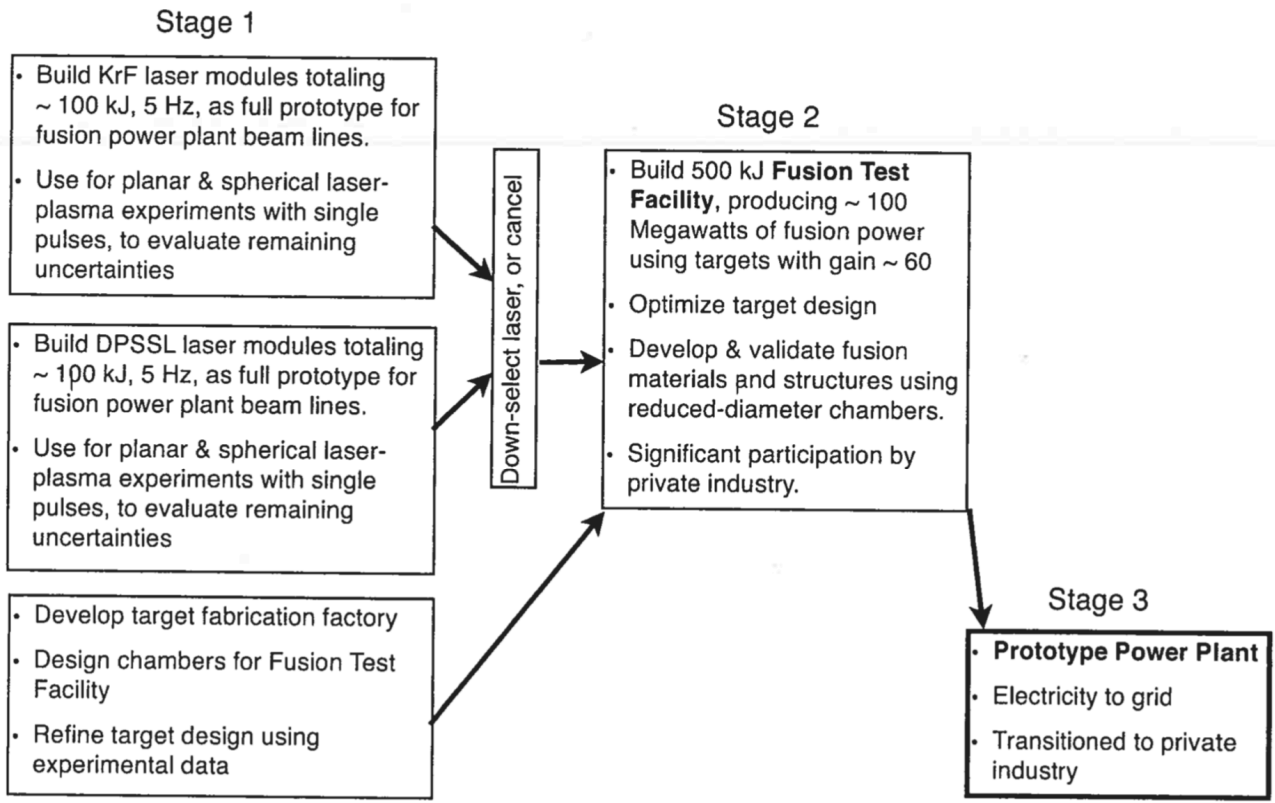

Figure (14): Steven Bodner's proposed plan for laser development for laser fusion.

While this author basically agrees with Bodner's plan there are several areas where he does not. Let's consider first Stage 1. In the intervening years, NRL has shifted its emphasis to ArF lasers. These radiate at 193 nm wavelength instead of the KrF laser's 248 nm, giving significant advantages to the laser target interaction. On a shoe string budget, NRL has shown that these operate well at the two hundred Joule level, while their KrF lasers have long been demonstrated to operate well at the multi kilo Joule level. [89]

Secondly, Bodner's plan, at the transition from stage one to stage two has some tough transition criteria, too tough in this author's opinion. First of all, say ArF reaches that milestone first. Does that really mean that DPSSL lasers should just be immediately abandoned, or visa versa? Perhaps further research, at lower level, by the temporary loser could still develop to a superior product. Secondly if neither of the lasers reach the required level for transition to stage 2 by some deadline, should the whole project be abandoned?

To this author, neither of these hard end points make any sense. The entire history of the fusion projects is one of both substantial progress, but coupled with large cost overruns and delays. Is it really reasonable to demand that these suddenly stop now, and expect all progress to be made strictly according to the initial budget and time line, or else? This author thinks not. He thinks that the only reason for such an abrupt end for laser fusion should be if the tokamak project forges so far ahead that pursuing laser fusion no longer makes sense.

As argued in Section II, this energy development program is of extreme importance for the continuation of modern civilization. While obviously every effort should be made to develop the many parts of it as rapidly and as economically as possible, it should not be held strictly to any schedule or budget. If it takes and extra decade, or costs a couple of $$B more, it is not the end of the world. It is a very difficult, but vitally important project. It is not easy to predict its ultimate cost or schedule. If it is delayed, there will still be plenty of fossil and nuclear fuel to get us through a decade or two.

The third leg of the laser development plan is obvious. Since LLNL has made such an important leap, and has a mega Joule laser right now, its work obviously should be continued. Also it has a huge infrastructure in place, one which would not be possible to rapidly duplicate anywhere else. Look at the photo of the participants of HAPL meetings. There were ~ 50-100 people. Then look at the picture LLNL NIF group shown in their September 2021 zoom seminars. There were ~ 500-1000 people in the picture. They are certainly the giant, the elephant in the room. If LLNL gets to the point where it is able to produce burning plasmas frequently and reliably, it could have a big impact whether the ultimate machine turns out to be a tokamak or laser. For instance it could begin experimental studies of breeding tritium and $^{233}$U, and also begin studies of recapturing unburned tritium.

With its Q = 0.67 result, LLNL is obviously on a roll, it should continue along the present track, but not for that much longer. Probably in a year or two, it will achieve a larger Q, maybe even 2 or 3, and if they achieve this, once again it will be a spectacular achievement. However where the goal becomes energy, it has no choice ultimately, but to shift away from indirect drive to direct drive. Indirect drive, just puts too much laser energy everywhere but the target, see Fig (8).

But first let us briefly discuss an intermediate option. The URLLE group has done a considerable amount of work on what they call polar direct drive. [90] This uses the optics basically as they are, but modified for direct drive, so that the illumination is mostly at the poles and is not spherically uniform. URLLE thinks that it can partially make up for the lack of spherical illumination in other ways. Both Bodner and this author are skeptical of polar direct drive. Here is Bodner:

Their polar direct drive proposal is feasible, but there are uncertainties and it is premature to evaluate their chance of success.

Here is the author from his 2014 paper [34]

This author worries that polar direct drive could be a large time and dollar sink spent on a non-optimum configuration.

This author believes a viable alternative is to convert NIF to direct drive illumination. Bodner argues against this for several reasons. He first states

Unofficial and rumored estimates from LLNL say that the conversion to symmetric illumination for direct drive would cost over $300 M and take at least 2 years. Since the paying customer is the weapons program, it won't happen.

Then he also argues that the laser is not good for direct drive for a variety of reasons, including insufficient bandwidth, non-optimum optical smoothing, etc. As Bodner put it:

No one in the direct drive program would have voluntarily chosen a NIF-type laser to test their target designs.

To consider these objections one by one, first of all, priorities do change in the course of a project. When NIF was built, the priority was stockpile stewardship. Now it should certainly be energy; some things will have to change. Furthermore, the weapons program will have had the laser in their desired configuration for 15 years. Now it time for a change to a direct drive configuration, which is optimal for energy; and anyway, who says that even in a direct drive configuration NIF cannot still have important applications for the weapons program.

Bodner's phase one would use lasers that are of the type much more optimal for laser fusion, but they would be 'only' 100 kJ. But if we have another mega Joule laser just sitting around, shouldn't we use it, even if it is not ideal? Bodner apparently thinks not. To this author it is a no brainer. Furthermore, URLLE believes that they can get decent results using a polar direct drive configuration with NIF. Surely, they must believe they can do better with uniform $4\pi$ illumination. Fig (8) shows, Livermore achieved is burning plasma with only ~ 10% of the laser energy absorbed by the target. Surely it must think that getting 80-90% on target (see Fig 9) might well improve things. Even a very sub optimal implosion from NIF in a direct drive configuration, would teach us great deal about burning plasmas, information that could be useful not only for the laser program, but also for ITER.

Note that Fig (10) shows that with direct drive, large gains are available. As in the author's 2014 paper [red, for hindsight, is added today]:

"Also, direct drive gain calculations show impressive gains at half a megajoule laser energy. NIF has nearly 4 times this. Hence there is a very large margin for error both regarding the laser energy and the gain calculations. Let's say NIF does a symmetric direct drive experiment and gets a gain of 'only' 10. (with the benefit of hindsight, I would have said 'only' 2 or 3.) Wouldn't this be a tremendous accomplishment? It might be just 2 or 3 years away."

Here is Bodner on how to design the target and laser:

Design both the fusion target and the laser with significant safety features. Then if there are surprises, one can recover.

These seems to say that even with a non-ideal 1 MJ laser, there is a very significant amount that can be learned from implosion experiments with NIF in a direct drive configuration. These 3 legs for the laser development seem to this author to be a reasonable plan for developing fusion breeding via laser fusion by midcentury.

But now let us consider what comes next. Bodner's phase 2 is the use of a 500 kJ laser to achieve a gain of 60. Then the goal would be to go on to phase 3 with a much higher gain and higher laser efficiency.

But to achieve commercial fusion breeding, a somewhat modified phase 2 might be all that is needed. There might well be no need for his phase 3, if fusion breeding is the goal. The laser with a gain of 60 and an efficiency of 5% would almost certainly be fine for laser fusion breeding.

So instead of a 500 kJ laser for phase 2, let us consider a 1 MJ laser which satisfies all of Bodner's phase 2's requirements. This seems reasonable, after all, we have a 1 MJ laser right now, but we do not have a target design which we can say with any confidence will have a gain of anywhere near 60. This would (hopefully) be accomplished in phase 2. So, at the end of phase 2, we would expect to have a 1 MJ laser and a target with a gain of 60 to produced 60 MJ of output fusion energy. However, used as a breeder, this 60 MJ of fusion neutron energy means 600 MJ of fuel, plus another 60 MJ of energy from the breeding of $^{233}$U. Now say this runs with a 20 Hz rep rate. This means it would produce 12,000 MW of nuclear fuel, enough to fuel four 1 GWe LWR's and also supply 400 MWe to the grid.

In other words, using fusion breeding as a goal, rather than pure fusion, the laser fusion program can skip Bodner's phase 3, and use a successful, modified phase 2 to go right to commercial power and fuel production.

VII. What if a viable pure fusion device suddenly becomes available?

Let's say that somehow a viable pure fusion device becomes available, let's say a tokamak with a 3-meter major radius, which produces 3 GWth of 14 MeV neutrons. This obviously violates CDR's by a large margin. Let's also stipulate that it needs only 30 MW of input power, small enough that we, so we can forget about it in our calculations. Then run through a conventional heat exchanger, it produces 1 GWe. Let's say the device only costs $4B and lasts for 30 years, or costs ~ $130M per year. Let's say that interest is another $130M, recapitalization over the 30 years costs another $100M, and operating expenses are $500M per year for a total cost of $860M per year. This means it costs $0.1 per kWhr to produce, so let's say it sells the power for 12 cents per kWhr. This could be a successful use for fusion.

Is this the end of the story for fusion breeding? Probably not! First of all, such a tokamak is most likely not possible. However, let us see what this it can do as a fusion breeder. Here we consider a blanket not mainly to exchange the heat generated by the 14 MeV neutrons, but to maximize additional neutron production. First breed the tritium using the $^7$Li reaction. This takes about 3 MeV from the fusion neutron, but does preserve a neutron. The fusion neutron now has ~ 11 MeV. Then consider a second blanket of mostly beryllium. It can generate additional neutrons at an energy cost of ~ 3 MeV per neutron. Thus we may be able to generate as many as 3 additional neutrons from the 11 MeV neutron after it has produced the tritium atom, making 4 neutrons all together. Let's say that half of these neutrons are used to generate $^{233}$U atoms and the half are lost due to various loss mechanisms. Hence each 14 MeV neutron could ultimately generate two $^{233}$U fuel atoms producing a total of ~ 400 MeV when burned. This reactor with the enhanced breeding blanket, could fuel as many as 25-30 one GWe reactors at a fuel cost of well under a penny per kWhr, almost certainly cheaper than mined uranium will be then.

So how would society use this reactor; as a power source, or as a breeder? Clearly, it is impossible to know, this not a decision for us to make, but for our children and grandchildren. This author's guess, but only a guess, is that a breeder would be a better choice. The point is that planning for a breeder now, with reactors we may well have a decent idea how to build, does no harm, but does only good. If now we can foresee only such breeders, as is the reality today, they will be a huge benefit to society. However, if by some miracle, we see how to build an economic pure fusion reactor, it will be an even better, cheaper breeder. Society will be able to make a choice as to which path it would like to follow.

VII.   The energy park

The Energy Park

Since 2004, every article the author has written on fusion breeding has ended with a section on 'The Energy Park', and this one is no different. The energy park is a proposed key element of an energy infrastructure to supply tens of terawatts to the world. It uses the fact that a fusion breeder can breed fuel for at least 5 LWR's of equal power, and each year an LWR discharges about 1/5 of its fuel as plutonium and higher actinides. [43]

The energy park proposes to burn the discharged actinides with a fast neutron reactor like the IFR. This is different from the French approach, which recycles these actinides to fuel for thermal reactors. But the thermal reactor also creates additional actinides, of constantly higher atomic number, and as the process is repeated, produces a more and more complicated stew of nuclear wastes. The advantage of fast neutron reactor is that one time through, it burns all actinides equally as the cross sections shown in Section V have shown. There is no endless recycling, a single burn will take care of all the actinides.

This series of papers have invariably assumed that the waste products of the thermal reactors must be rendered harmless. The alternative is burying them somewhere, and creating what amounts to a 'plutonium mine', which will plague society for half a million years or so. This is an immoral burden to lay upon our descendants, hence the need for the fast reactor in the energy park.

One envisions an energy infrastructure where there is one fusion breeder to supply fuel to about 5 thermal reactors like the LWR or CANDU or more advanced reactor, and one fast neutron reactor to burn the 'waste' actinides.

The fast neutron reactor could be something like the Integral Fast Reactor (IFR), developed by Argonne National Laboratory. It ran successfully at 60 MW for years before it was disassembled. It could run on any actinide and could run in either a breeder or burner mode. As we see in Fig (2), at ~1-2 MeV neutron energy, fissile and fertile materials have about the same fission cross sections. Thus, the IFR can be run in a mode to simply 'burn' any actinide. Specifically, it could be used to burn all the plutonium and other higher actinides that an LWR discharges.

The British, who have the largest plutonium 'waste' stockpile, are now seriously considering constructing a much larger version of the IFR called PRISM to 'treat' their large stockpile of plutonium waste. Perhaps they are making an important step in the ultimate development of the energy park.

A schematic of the energy park is shown in Figure (15). Most of the elements of the energy park are available today, only the fusion breeder needs full development.

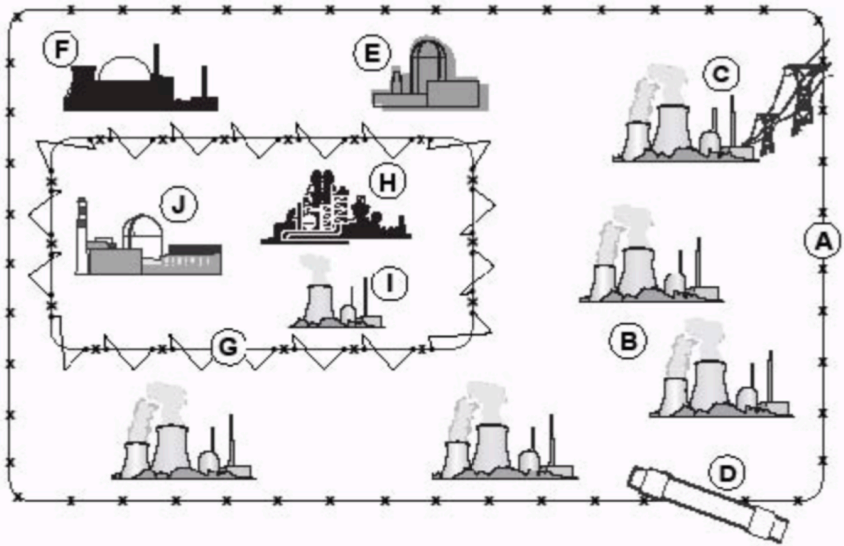

Figure 15: The energy park: A. low security fence; B. 5 thermal 1GWe nuclear reactors, LWRs or more advanced reactors; C. output electricity; D. manufactured fuel pipeline, E. cooling pool for storage of highly radioactive fission products for 300–500 years necessary for them to become inert. This is a time human society can reasonably plan for, unlike the ~ half million years it would take for the plutonium 'waste' to be buried in a repository, essentially creating a plutonium mine; F. liquid or gaseous fuel factory; G. high security fence, everything with proliferation risk, during the short time before it is diluted or burned, is behind this high security fence; H. separation plant. This separates the material discharged from the reactors (B) into fission products and transuranic elements. Fission products which have commercial value would be separated out and sold, the rest go to storage (E), transuranic elements go to (I); the 1GWe integral fast reactor (IFR) or other fast neutron reactor where actinides like plutonium are burned; J. the fusion breeder, producing 1GWe itself and also producing the fuel (ultimately enriched to ~4% $^{233}$U in $^{238}$U) for the 5 thermal nuclear reactors for a total of 7 GWe produced in the energy park.

The world-wide use of energy parks could generate carbon free power, in an economically and environmentally viable way, and with little or no proliferation risk. They could supply tens of TW at least as far into the future as the dawn of civilization was in the past.

IX Conclusions

To achieve economic power by magnetic fusion energy (MFE), once ITER is successful, one has to proceed to the next step, the DEMO as the ITER web site says. To proceed via inertial (i.e. laser) fusion energy (IFE), there is less information on a plan to do so, but Steven Bodner, former head to the NRL program suggested a 3 step process. These last steps for each path would be very large steps, most likely taking years or decades, and costing tens of billions.

This paper asserts that with fusion breeding instead of pure fusion, one can skip the DEMO if one goes via the MFE route, and can skip Bodner's final stage if one goes the IFE route. Perhaps, finally (!), fusion power really can be achievable in 35 years; and not available in 35 years as it always will be.

Fusion research is extremely important for the maintenance of civilization. It will not be achieved quickly, no matter how much it claims to be a quick solution to a nonexistent 'climate emergency'. There is no way to avoid the reality that either fusion breeding, and especially pure fusion will become available only with a huge effort which will take several decades and will cost billions. But it is more than worth the investment. The continuation of modern civilization could well depend on its success.

Acknowledgement: This paper was not supported by any organization public or private. This paper is dedicated to the memory of Professor David Rose of MIT. He was the leader of the MIT fusion program when I was a graduate student and junior faculty member there. He always expressed confidence that I would make a decent contribution to fusion. I hope my efforts up to now, and this work especially have proven his prediction to be correct.

20. JUDITH CURRY blog, https://judithcurry.com/  She is the former head of the earth science department at Georgia Tech, but quit when the academic environment got too stultifying and restrictive for her.

21. WALLACE MANHEIMER, *Original sin, prophets, witches, communists, preschool sex abuse, and climate change*, International Journal of Engineering and Applied Sciences (IJEAS) ISSN: 2394-3661, Volume-4, Issue-7, July 2017

https://www.ijeas.org/download_data/IJEAS0407025.pdf

22. WALLACE MANHEIMER, *Climate change, on media perceptions and misperceptions*, Forum of Physics and Society, (American Physical Societal essay journal) October, 2019
https://www.aps.org/units/fps/newsletters/201910/climate-change.cfm

23. SEAVER WANG and ZEKE HAUSFATHER, *Climate Change: Robust Evidence of Causes and Impacts*, Forum on Physics and Society, January 2020
https://engage.aps.org/fps/resources/newsletters/january-2020

24. WALLACE MANHEIMER, *Media, Politics and Climate Change, a response to Wang and Hausfather*, Forum on Physics and Society, April 2020
https://www.aps.org/units/fps/newsletters/202004/media.cfm

25. WALLACE MANHEIMER, *Some dilemmas of climate simulations* wattsupwiththat, April 27, 2020
https://wattsupwiththat.com/2020/04/27/some-dilemmas-of-climate-simulations/

26. JOHN R. CHRISTY of University of Alabama Huntsville, U.S. House Committee on Science, Space & Technology 2 Feb 2016 Testimony of University of Alabama in Huntsville.
https://docs.house.gov/meetings/SY/SY00/20160202/104399/HHRG-114-SY00-Wstate-ChristyJ-20160202.pdf

27. PAUL VOSSEN, *Climate scientists open up their black boxes to scrutiny*, Science, Vol 354, issue 6311, page 401, October 26, 2016
56

90. R.S. CRAXTON et al, Polar direct drive: Proof-of-principle experiments on OMEGA and prospects for ignition on the National Ignition Facility, Physics of Plasmas **12**, 056304 (2005)